\documentclass[%
reprint, onecolumn,
amsmath,amssymb,
aps,
]{revtex4}

\usepackage{graphicx}
\usepackage{dcolumn}
\usepackage{bm}
\usepackage{comment}
\usepackage{caption}
\usepackage{subcaption}
\usepackage{float}
\usepackage{amsmath}
\usepackage{mathtools}
\usepackage{hyperref}
\usepackage[Document]{ragged2e}

\newcommand{\upperRomannumeral}[1]{\uppercase\expandafter{\romannumeral#1}}

\begin{document}
	\newcommand{\be}{\begin{equation}}
		\newcommand{\ee}{\end{equation}}
	\newcommand{\bea}{\begin{eqnarray}}
		\newcommand{\eea}{\end{eqnarray}}
	\newcommand{\RN}[1]{\textup{\uppercase\expandafter{\romannumeral#1}}}
	\preprint{APS/123-QED}

	\title{{Optimum transport in systems with time-dependent drive and short-ranged interactions}}

	\author{Deepsikha Das}

	\author{Punyabrata Pradhan}
	\email{punyabrata.pradhan@bose.res.in}

	\author{Sakuntala Chatterjee}
	\email{sakuntala.chatterjee@bose.res.in}
        
	\affiliation{Physics of Complex Systems, S.N. Bose National Centre for Basic Sciences \\
	Block-JD, Sector-III, Salt Lake, Kolkata 700106, India}

	\begin{abstract}
	
We study one-dimensional hardcore lattice gases, with nearest-neighbor interactions, in the presence of an external potential barrier, that moves on the periodic lattice with a constant speed. We investigate how the nature of the interaction (attractive or repulsive) affects particle transport and determine, using numerical simulations and mean-field calculations, the conditions for an optimum transport in the system. Physically, the particle current induced by the time-dependent potential is opposed by a diffusive current generated by the density inhomogeneity (a traveling wave) built up in the system, resulting in a current reversal, that crucially depends on the speed of the barrier and particle-number density. Indeed the presence of nearest-neighbor interaction has a significant impact on the current: Repulsive interaction enhances the current, whereas attractive interaction suppresses it considerably. 
 Quite remarkably, when the number density is low, the current increases with the strength of the repulsive interaction and the maximum current is obtained for the strongest possible repulsion strength, i.e., for the nearest-neighbor exclusion.
 However, at high density, very strong repulsion makes particle movement difficult in an overcrowded environment and, in that case, the maximal current is achieved for  weaker repulsive interaction strength.

	\end{abstract}
	
	\maketitle
	
	
	\section{Introduction}

The ability to manipulate colloidal-particle motion in narrow channels using time-varying optical potential has opened up new research avenues in driven diffusive systems  \cite{1, 16, 18, 100, 13}. 	
These studies have provided valuable insights into a variety of important aspects of non equilibrium systems, such as verification and applicability of fluctuation relations, among other things.
Experiments with colloidal particles driven by an optical trap \cite{14, 15} have previously established the validity of the fluctuation-dissipation theorem, which predicts entropy production over finite time, and that of a generalized Einstein relation \cite{12}; Additionally, the violation of the second law of thermodynamics has been experimentally demonstrated for small systems over short time scales \cite{10}; see Ref. \cite{21} for review. 
Recently, a particularly promising research direction that has received a lot of attention is the characterization of particle transport in a periodically driven many-particle system \cite{21, 23}. These systems find application in a wide range of situations. For example, stochastic pumps \cite{19, 26, 27, 29}, in which the time-varying external parameters drives the systems away from equilibrium, can generate a directed particle flow; also consider the thermal ratchets \cite{20, 53, 30}, where non equilibrium fluctuations can induce a directed particle motion. Indeed, much attention has been focused on, and significant progress has been made, in understanding the underlying mechanism of directed flow in thermal ratchets and molecular pumps \cite{19}.

Notably, the characterization of particle transport in time-varying external potential is important also in the context of driven fluids in confined geometry, leading to the identification of number of unexpected consequences, such as negative differential resistance and absolute negative mobility, etc., among others  \cite{5, 6, 2, 61, 62, 64, 66, 67}.
In the past, particle transport in colloidal suspensions in narrow channels have motivated studies of noninteracting particles, driven by a moving potential barrier, using dynamic density functional theory \cite{5, 6}.
Eventually, several many-particle models were also put forth in an effort to theoretically understand the role of hardcore interactions in these systems \cite{31}. One particularly important question in this context is whether the system can support a nonzero dc (time-averaged) current when it is driven by a time-periodic driving force. Although the presence of an external forcing would typically suggest the presence of a current in the system, the periodic nature of the driving however means that the net force acting on the system over a time period is zero. In that case, do such systems still carry a nonzero dc current? If so, in what direction does the current flow?
Another intriguing question is whether it is possible to optimize the particle current by tuning various control parameters.

In order to address the above issues, a series of works \cite{31, 32, 35} considered the paradigmatic models of simple exclusion processes \cite{37}, in which the interaction among the particles was assumed to be of the simplest possible form, i.e., hardcore exclusion. The motion of the particles were described on a lattice where particles hop from one site to a neighboring unoccupied site; in that case, the periodic external potential was simply represented by space and time dependent hopping rates.
Depending on whether the time-varying hopping rates were present only on particular sites or were present throughout the system, it was shown through numerical simulations and a perturbative approach that the dc current flowing through such a system could either vanish (inversely with system size) or have a finite value. 
Furthermore, the dc current was found to exhibit non-monotonic dependence on the time-period of the drive. Several interesting features such as current reversal and system-size dependent transport were observed \cite{35} in the case when time-varying potential maintains a position-dependent phase relation among sites, that results in a nonzero dc current.

Subsequently, in another study of a many-particle lattice model \cite{33}, our group had developed a new and simple method of modeling a periodically moving drive in a system of hardcore particles diffusing on a ring. Motivated by moving trap or barrier used in experiments with particles in an optical potential, we studied a system with a  ``defect'' site, which had a hopping rate different from the rest of the system. Then, in the non equilibrium setting, the defect site was considered to move on the lattice with a speed $v$, and, for a one dimensional ring of $L$ sites, to complete one cycle  after a time period $L/v$. Using numerical simulation and a mean-field theory, we observed that, in the time-periodic steady state, a density inhomogeneity is created around the defect, resulting in a dc current in the system that scales as $1/L$. The direction and magnitude of the dc current was controlled by tuning the defect speed, particle density and the bulk diffusivity of particles. Moreover, in the presence of multiple defect sites \cite{34}, an interesting collective behavior was observed when the defect sites were close enough so that their respective density patterns generated by each of the defects overlap with each other.
Interestingly, reversal of current has also been observed in a slightly different set up \cite{300} in the context of a single particle, which  diffuses in a two-dimensional channel of varying width and is driven by a force having a random orientation across the channel; in this case, the current reversal happens by tuning both the transverse and the longitudinal drive.

So far, in the previous studies of many-particle lattice models, the only type of the interaction considered between the particles was the hardcore exclusion. However, in real systems, particles can also experience short-ranged attraction or repulsion and the interplay between external driving and inter-particle interactions are expected to give rise to nontrivial effects. In order to investigate this scenario, in the present work we consider a many-particle lattice model in which hardcore particles diffuse and interact via nearest-neighbor attractive or repulsive potential. In other words, in addition to the hardcore exclusion, a particle, in the case of repulsive (attractive) interaction, now prefers to have its neighboring site empty (occupied). Here we are primarily interested in exploring how the strength of the interaction potential affects the particle current in the system.  Does the system still supports current reversal and, if so, how are the transport characteristics affected by many-particle interactions? Is there an optimum interaction strength for which magnitude of the current in either direction is largest?

In this paper, by performing Monte Carlo simulations and using a modified mean-field theory, we have determined the condition of optimal transport in the system and studied how attractive or repulsive interaction among the particles affect the transport. We show that a moving defect always induces current in the negative direction, i.e., along the direction opposite to the defect movement. But, due to the density inhomogeneity produced by the defect movement, the diffusive current in the system flows in the positive direction. As a result, when the bulk diffusion in the system is negligibly small, we find current in the negative direction. However, as the bulk diffusion becomes stronger, the current changes sign and becomes positive. By varying the defect speed, particle density and the interaction strength, we determine the parameter regime, that yields the optimum current in the system in either direction. It turns out that an attractive interaction among the particles hinders transport, while a repulsive interaction enhances it. For small particle density, current is largest when the strength of the repulsive interaction assumes its highest possible value. However, for large particle density, the system is overcrowded and a very strong repulsion indeed blocks certain transitions and consequently reduces the current. In that case, the optimum transport is obtained when the interaction strength lies somewhat below the largest possible value. However, unlike repulsive interaction and irrespective of defect speed and bulk density, the current decreases monotonically with the attractive interaction strength.


The organization of the paper is as follows: We describe the model in section II. In Section III, we describe the simplest case where dynamics in the bulk of the system is absent (the case where particle speed is larger compared to the corresponding rate of bulk diffusion). Analytical formalism for this particular situation is presented in subsections III A and III B while results are shown and discussed in subsection III C. In section IV, we discuss the case when bulk dynamics is also compared to the other rates in the system. Our conclusions are presented in section V.

	\section{The Model} \label{model}

	 We consider paradigmatic models of exclusion processes involving hardcore particles with nearest-neighbor interaction \cite{43}. We incorporate the periodically moving external potential barrier simply as a set of moving ``defects'' \cite{33, 34}, each of them resides at a site for duration $\tau$ before moving to the right. The energy function for the system can be written as
	\begin{equation}
		H = -\dfrac{J}{2}\sum_{\substack{i, j \\ <i, j>}}\eta^{\{\alpha_{k}\}}_{i} \eta^{\{\alpha_{k}\}}_{j} + \sum_{i}\eta^{\{\alpha_{k}\}}_{i} V_{i} 
		\label{hamiltonian}
    \end{equation}
	where, $\eta^{\{\alpha_{k}\}}_{i}$, $\eta^{\{\alpha_{k}\}}_{j}$ denote occupancy of sites which can take values 0 or 1. The indices $\{\alpha_{k}\} \equiv \{\alpha_{1}, \alpha_{2}, \dots ,\alpha_{N}\}$ are a set of $N$ elements with the $k$-th element, $\alpha_{k}$, denoting the position of the $k$-th defect and $<i, j>$ denotes that sites $i$ and $j$ are the nearest neighbors, A site is called a defect site when a potential barrier is present there and it is called a bulk site otherwise. The potential at site $i$ is $V_{i} = \sum_{k}V_{0} \delta_{i,\alpha_{k}}$, where $V_{0}$ represents height of the onsite potential barriers, and $J$ denotes the interaction strength that can vary in the range $-\infty$ to $+\infty$. 
	 A mapping \cite{44} can be performed from $J$ to a dimensionless parameter $\epsilon$, as given below
	\begin{equation}
		\begin{aligned}
			e^{-\beta J} = \dfrac{(1+\epsilon)}{(1-\epsilon)}
		\end{aligned}
		\label{mapping}
	\end{equation}
 such that $|\epsilon| \leq 1$ and thus $0 \le e^{-\beta J} < \infty$.	
Note that, for a left (right) hopping to take place, the departure site has to be occupied by a particle and its immediate left (right) site has to be empty. Therefore, particle hopping towards left (right) can happen in four possible ways and all possible transition rates \cite{44} for left and right hops are written as following: \\
	
	\begin{minipage}{0.5\textwidth}
		\centering
		\begin{equation*}
			\begin{array}{ccl}
				0010 \xrightleftharpoons[c e^{-\beta V_{i}}]{c} 0100 \\
				\parskip 10 mm
				1010 \xrightleftharpoons[c(1+\epsilon) e^{-\beta V_{i}}]{c(1-\epsilon)} 1100 \\
				\parskip 10 mm
				0011 \xrightleftharpoons[c(1-\epsilon) e^{-\beta V_{i}}]{c(1+\epsilon)} 0101 \\
				\parskip 10 mm
				1011 \xrightleftharpoons [c e^{-\beta V_{i}}]{c} 1101
			\end{array}
		\label{lefthop}
		\end{equation*}
	\end{minipage}%
	\begin{minipage}{0.5\textwidth}
		\centering
		\begin{equation}
			\begin{array}{ccl}
				0100 \xrightleftharpoons[c e^{-\beta V_{i}}]{c} 0010 \\
				\parskip 10 mm
				0101 \xrightleftharpoons[c(1+\epsilon) e^{-\beta V_{i}}]{c(1-\epsilon)} 0011 \\
				\parskip 10 mm
				1100 \xrightleftharpoons[c(1-\epsilon) e^{-\beta V_{i}}]{c(1+\epsilon)} 1010 \\
				\parskip 10 mm
				1101 \xrightleftharpoons[c e^{-\beta V_{i}}]{c} 1011
			\end{array}
		\label{righthop}
		\end{equation}
	\end{minipage}  

	\parskip 0.3 cm
	with $\beta V_{0} = \ln{(p/r)}$, where $\beta$ is the inverse of temperature and (i) $c=p/2$ when the departure site is a defect site and a particle hops out of it, (ii) $c=q/2$ when the departure site is a bulk site, (iii) $c=r/2$ when the departure site is a neighbor of a defect and the destination is a defect site. The transition rates and corresponding reverse rates follow detailed balance condition for defect velocity $v=0$. For simplicity, in our study, we have considered only a single defect, which represents an infinite potential barrier, periodically moving over a lattice of length $L$ with speed $v$. Consequently $r=0$ and $p=1$ are maintained throughout the paper.

	\begin{figure}[H]
		\centering
		\includegraphics[scale=0.58]{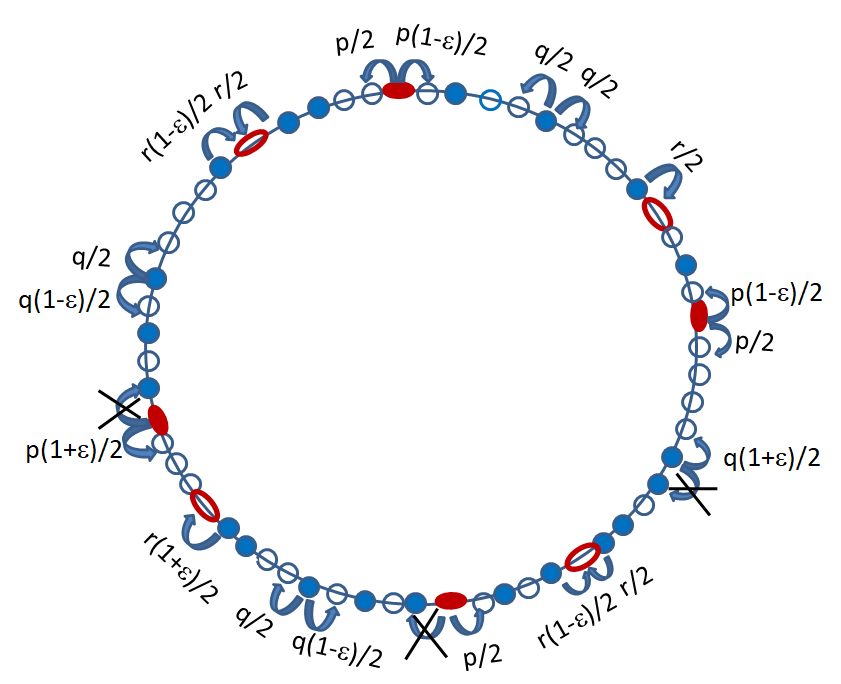}
		\caption{Schematic diagram of the model. Red solid (empty) ellipses represent occupied (empty) defect sites while the blue solid (empty) circles are occupied (empty) bulk sites. In one 
		time step a particle can jump to one of its neighboring site provided the destination site is empty. The transition rates depend on the local configurations around the departure 
		site, as specified in Eq. \eqref{righthop}.}
	\label{TheModelFig}
	\end{figure}
	

	\section{\boldmath$p=1$, $r=0$, $q=0$: No dynamics in the bulk} \label{q0}
        
	\subsection{Analytical formalism for a periodically moving defect} \label{analytics}
	
	In this section, we consider the simplest case when the inverse hopping rate is much larger compared to the typical residence time scale of the defect. In that case, we assume the bulk hopping rate $q=0$, i.e., dynamics in the bulk is completely frozen. A particle can hop during an infinitesimal time interval $dt$ only if its position coincides with the position of the defect denoted by $\alpha$. Starting from an initial configuration, the system reaches a time periodic steady state, given sufficient time has passed. The density profile of the system has a form of a traveling wave moving over the lattice with the same speed $v$ as that of the moving defect. The defect spends time $\tau$ at a particular site before moving on to the next site where $\tau = 1/v$ is the residence time of the defect. When the defect spends one Monte-Carlo step on each lattice site, $v$ is measured as $1$. We have measured density profile at time-steps $t=n\tau$ just before the defect moves on to the next site, after spending time $\tau$ at the previous site, with $n = 0,1,2,...., \infty$. For time $t=n\tau$ we write the discrete time evolution equation \cite{33} for density $\rho_{i}^{(\alpha)}(t) = \langle\eta_{i}^{(\alpha)}(t)\rangle$,
	\begin{equation}
		\langle\rho^{(\alpha + 1)}(t + \tau)|=\langle\rho^{(\alpha)}(t)|W^{(\alpha + 1)}
		\label{timeevolution}
	\end{equation}
	Here, $\langle\rho^{(\alpha)}(t))| \equiv \{\rho_{1}^{(\alpha)}(t),....,\rho_{i}^{(\alpha)}(t),....,\rho_{L}^{(\alpha)}(t)\}$ is a row vector of length $L$, its $i$-th element being $\rho_{i}^{(\alpha)}(t)$ with $\alpha$ denoting the position of the defect. \( W^{(\alpha + 1)} \) is the transition matrix with the defect site at $\alpha$ + 1. Its structure depends on the position of the defect site. For example when $\alpha + 1 = 1, 2$, respective transition matrices can be written as, \\
	\begin{minipage}{0.5 \textwidth}
		\begin{equation*}
			W^{(1)} = 
			\begin{bmatrix}
				1-a_{+}-a_{-} & a_{+} & 0 & \dots & 0 & a_{-} \\
				0       &  1    & 0  & 0 & \dots & 0 \\
				\dots   & \dots & \dots & \dots & \dots & \dots \\
				\dots   & \dots & \dots & \dots & \dots & \dots \\
				0       & \dots & 0  & 0 & 1     & 0    \\
				0       & 0     & \dots  & 0   & 0     & 1 
			\end{bmatrix}
		\label{W1}
		\end{equation*}
	\end{minipage}%
	\hspace {0.01 cm}
	\begin{minipage}{0.5 \textwidth}
		\begin{equation}
			W^{(2)} = 
			\begin{bmatrix}
				1 & 0 & 0 & \dots & 0 & 0 \\
				a_{-} & (1-a_{+}-a_{-}) & a_{+} & 0 & \dots & 0 \\
				0   & 0 & 1 & 0 & \dots & 0 \\  
				\dots   & \dots & \dots & \dots & \dots & \dots \\
				0 & \dots & 0 & 0 & 1 & 0 \\
				0 & 0 & \dots & 0 & 0 & 1 
			\end{bmatrix}
		\label{W2}
		\end{equation}
	\end{minipage}
	
	The \(i,j\)-th element of the transition matrix \cite{33} can be written as,
	
	\begin{align} \label{Wm}
		W_{ij}^{\alpha + 1} &= 1-a_{+}-a_{-} \quad \quad \text{for} \quad  \quad i=j=\alpha + 1 \nonumber \\
		W_{ij}^{\alpha + 1} &= a_{-} \quad      \quad  \text{for}  \quad \quad i=j+1=\alpha + 1 \nonumber   \\
		W_{ij}^{\alpha + 1} &= a_{+}  \quad     \quad  \text{for} \quad \quad i=j-1=\alpha + 1    \\
		W_{ij}^{\alpha + 1} &= 1      \quad     \quad  \text{for} \quad \quad i=j \neq \alpha + 1 \nonumber \\
		W_{ij}^{\alpha + 1} &= 0   \quad   \quad  \text{for} \quad \quad i \neq \alpha + 1,\,i \neq j  \nonumber 
    \end{align}

	Here $a_{\pm}$ are the conditional probabilities that, given the defect site is occupied, particle from the defect site moves to its unoccupied right(left) neighboring site during the residence time $\tau$. Starting from microscopic dynamics we can have their expressions as following:
	
	\begin{align} \label{apmexact}
		a_{+} = \sum_{m=1}^{6} {\mathcal C}_{m}^{+} \omega_{m}^{+} \quad , \quad & a_{-} = \sum_{n=1}^{6} {\mathcal C}_{n}^{-} \omega_{n}^{-} 
	\end{align}

	where ${\mathcal C}_{m}^{+}$, ${\mathcal C}_{n}^{-}$ are the conditional probabilities of different local configurations favorable for right and left hopping respectively during the residence time $\tau$, given the defect site is occupied. $\omega_{m}^{+}$, $\omega_{n}^{-}$ denote transition probabilities for right and left hopping respectively from an occupied defect site during $\tau$. For example, 
	\begin{equation}
		{\mathcal C}_{1}^{+} = \text{Prob}.(00\hat{1}01 | \eta^{(\alpha)}_{\alpha + 1}=1) = \dfrac{\bigg\langle(1-\eta^{(\alpha)}_{\alpha - 1})(1-\eta^{(\alpha)}_{\alpha})\eta^{(\alpha)}_{\alpha + 1}(1-\eta^{(\alpha)}_{\alpha + 2})\eta^{(\alpha)}_{\alpha + 3}\bigg\rangle}{\langle\eta^{(\alpha)}_{\alpha + 1}\rangle} 
		\label{C1}
	\end{equation}
	
	\begin{equation}
		{\mathcal C}_{1}^{-} = \text{Prob}.(10\hat{1}00 | \eta^{(\alpha)}_{\alpha + 1}=1) = \dfrac{\bigg\langle\eta^{(\alpha)}_{\alpha - 1}(1-\eta^{(\alpha)}_{\alpha})\eta^{(\alpha)}_{\alpha + 1}(1-\eta^{(\alpha)}_{\alpha + 2})(1-\eta^{(\alpha)}_{\alpha + 3})\bigg\rangle}{\langle\eta^{(\alpha)}_{\alpha + 1}\rangle}
		\label{C2}
	\end{equation}
	
	\begin{equation}
		\omega_{1}^{+}   =   \omega_{1}^{-}  =  \dfrac{1-\epsilon}{2-\epsilon}\biggl(1-e^{-(2-\epsilon)/4v}\biggr)
		\label{omega}  
	\end{equation}
	where ${\mathcal C}_{1}^{+}$ and ${\mathcal C}_{1}^{-}$ represent the conditional probabilities for local configurations $00\hat{1}01$ and $10\hat{1}00$ respectively given the defect site is occupied. $\omega_{1}^{+}$ and $\omega_{1}^{-}$ denote the transition probabilities corresponding to configuration $00\hat{1}01$ and $10\hat{1}00$ respectively during a residence time $\tau$ with $\hat{1}$($\hat{0}$) denoting an occupied (unoccupied) defect site (details are given in Appendix). Due to the time periodic structure of the steady state, the density profile comes back to itself each time the defect moves across the ring and completes a cycle. So, $W^{(\alpha+1)}...W^{(L)}W^{(1)}...W^{(\alpha-1)}W^{(\alpha)}$ has an eigenvector $\langle\rho_{st}^{(\alpha)}|$, with eigenvalue unity. The steady state density \cite{33} at $i$-th site satisfies,
	\begin{equation}
		\rho_{st,i}^{(\alpha+1)} = \rho_{st,i-1}^{(\alpha)}
		\label{steadystatedensity}
	\end{equation}
	which follows from the time periodic structure of the steady state density and from Eq. \eqref{timeevolution}. To solve for the density profile in a time periodic steady state, we find that at the time of measurement, the defect site $\alpha$ registers lower density compared to the bulk as for $r$=0 particles can't hop into the defect site. Rather they can only hop out of the defect site. For $q$=0 the neighboring sites \( (\alpha \pm 1) \) can only receive particles from the defect site without any loss. The site $(\alpha+1)$ thus has a higher density compared to that at the bulk. On the other hand, the site $(\alpha-1)$ which was previously occupied by the defect and has already registered lower density, could only receive particle from the defect site $\alpha$ and its density goes back to the bulk level. Therefore regarding the structure of the density profile as a function of position, we formulate an ansatz \cite{33} in the form of a traveling density wave which moves with the defect $\alpha$.
	\begin{align}
		\rho_{st,i}^{(\alpha)} =  &  \rho_{-} \quad \text{for} \quad i =  \alpha \nonumber \\
		\rho_{st,i}^{(\alpha)} = &\rho_{+} \quad \text{for} \quad i=\alpha + 1 \\
		\rho_{st,i}^{(\alpha)} = &\rho \quad \quad \text{otherwise} \nonumber
	\label{rhopmrho}
    \end{align}
	The ansatz can be used in Eqs. \ref{timeevolution} and \ref{steadystatedensity}, to obtain Eqs. \ref{rhop}, \ref{rhom}.
	\begin{equation}
		\rho_{+}a_{-} + \rho_{-} = \rho_{b}
		\label{rhop}
	\end{equation}
	\begin{equation}
		\rho_{+}a_{+} + \rho_{b} = \rho_{+}
		\label{rhom}
	\end{equation}
	which can be solved by using particle number conservation \( \rho_{+} + \rho_{-} + (L-2)\rho_{b} = L\rho \) to get the exact densities,
	\begin{equation}
		\rho_{b} \quad = \quad \dfrac{(1-a_{+})L}{2-a_{+}-a_{-}+(1-a_{+})(L-2)}\rho \quad \simeq   \quad \rho
		\label{rho}
	\end{equation}
	\begin{align}
		\rho_{+} \quad \simeq \quad &\dfrac{1}{1-a_{+}}\rho 
		\label{rhopwrtapm}  \\
		\rho_{-} \quad \simeq \quad  &\dfrac{1-a_{+}-a_{-}}{1-a_{+}}\rho
		\label{rhomwrtapm}
	\end{align}
	as $L>>1$. From Eqs. \ref{rhopwrtapm} and \ref{rhomwrtapm} it is evident that \( \rho_{+} > \rho \) and \( \rho_{-} < \rho \) i.e a peak and a trough are formed in front of the defect site and at the defect site respectively.\\
	
	The trough and the peak present in the density profile are different in size which will result in a non-zero particle current in general. Contribution to particle current comes from only two bonds adjacent to the defect site as no hopping takes place across any other bond for $q=0$. Particle current consists of two components $\text{$J$}_{+}$ and $\text{$J$}_{-}$, defined to be the time rate of rightward and leftward movement of particles respectively, from the defect site. The total current is the algebraic sum of them. The defect visits a particular site with rate $v/L$. Thus the expression for particle current can be written as 
	\begin{equation}
		J = J_{+} + J_{-}
		= \dfrac{v}{L} (\langle\eta^{(\alpha)}_{\alpha+1}\rangle a_{+} - \langle\eta^{(\alpha)}_{\alpha+1}\rangle a_{-})
		\label{$J$expression1}
	\end{equation}
	\ which can be written in terms of $\rho_{\pm}$ from Eq. \ref{rhopwrtapm} and \ref{rhomwrtapm} as
	\begin{equation} 
		\begin{aligned}
			J = \dfrac{v}{L}((\rho_{+} - \rho) + (\rho_{-} - \rho)) = \dfrac{v}{L}(\rho_{+} + \rho_{-} - 2\rho)
		\end{aligned}
	\label{$J$expression2}
	\end{equation}
	where
	\begin{equation}
		J_{+} = \dfrac{v}{L} (\langle\eta^{(\alpha)}_{\alpha+1}\rangle a_{+}) = \dfrac{v}{L}(\rho_{+} - \rho)
		\label{$J$p}
	\end{equation}
	\begin{equation}
		J_{-} = -\dfrac{v}{L} \langle\eta^{\alpha}_{\alpha+1}\rangle a_{-} = \dfrac{v}{L}(\rho_{-} - \rho)
		\label{$J$m}
	\end{equation} 
	
	\subsection{Mean-field theory} \label{mean-field}
	
	The exact expression for $a_{+}$ and $a_{-}$ are given in Eq. \ref{apmexact} and in Eq. \ref{A1} to \ref{A40}. To write $a_{+}$ and $a_{-}$ as explicit functions of $\epsilon$, $\rho$ and $v$, we use mean-field approximation, where many-point correlations ${\mathcal C}_{m}^{+}$, ${\mathcal C}_{n}^{-}$ (for m, n = 1, 2,..., 6) are assumed to be factorized. Thus we have, 
	
	\begin{equation}
		\begin{aligned}
			{\mathcal C}_{1}^{+} = {\mathcal C}_{1}^{-} = \rho(1-\rho_{-})(1-\rho)^{2}
			\label{C1pm}
		\end{aligned}
	\end{equation}
	etc. and
	\begin{equation} \label{apfinal}
		\begin{aligned}
			a_{+} = (1-\rho)\Biggl[(1-\rho)(1-\rho_{-})\rho \omega_{1}^{+} 
			+ \rho^{2}(1-\rho_{-})\omega_{2}^{+} 
			+ (1-\rho)^{2}(1-\rho_{-}) \omega_{3}^{+} \\
			+ \rho(1-\rho_{-})(1-\rho) \omega_{4}^{+} 
			+ \rho_{-}(1-\rho) \omega_{5}^{+} 
			+ \rho_{-}\rho \omega_{6}^{+}\Biggr]
		\end{aligned}
	\end{equation}
	
	\begin{equation} \label{amfinal}
		\begin{aligned}
			a_{-}= (1-\rho_{-})\Biggl[\rho(1-\rho)^{2} \omega_{1}^{-} 
			+\rho^{2}(1-\rho) \omega_{2}^{-}  
			+(1-\rho)^{3}\omega_{3}^{-} \\
			+\rho(1-\rho)^{2} \omega_{4}^{-}
			+\rho(1-\rho) \omega_{5}^{-} 
			+\rho^{2} \omega_{6}^{-}\Biggr]
		\end{aligned}
	\end{equation}
	
	Combining Eqs. \ref{rhopwrtapm}, \ref{rhomwrtapm} we have the following form.
	\begin{equation}
		(\rho_{-}-\rho)(1-a_{+})+a_{-}\rho=0
		\label{intermediate}
	\end{equation}
	If expressions for $a_{+}$ and $a_{-}$ from Eqs. \ref{apfinal}, \ref{amfinal} are put into Eq. \ref{intermediate} we will obtain the following quadratic equation for $\rho_{-}$,
	\begin{equation}
		\begin{aligned}
			(\rho_{-}-\rho)[1-(1-\rho)\{(1-\rho)(1-\rho_{-})\rho \omega_{1}^{+} 
			+ \rho^{2}(1-\rho_{-})\omega_{2}^{+} 
			+ (1-\rho)^{2}(1-\rho_{-}) \omega_{3}^{+} \\
			+ \rho(1-\rho_{-})(1-\rho) \omega_{4}^{+} 
			+ \rho_{-}(1-\rho) \omega_{5}^{+} 
			+ \rho_{-}\rho \omega_{6}^{+}\}]+ \rho (1-\rho_{-})[\rho(1-\rho)^{2} \omega_{1}^{-} \\
			+\rho^{2}(1-\rho) \omega_{2}^{-} 
			+(1-\rho)^{3}\omega_{3}^{-} 
			+\rho(1-\rho)^{2} \omega_{4}^{-}
			+\rho(1-\rho) \omega_{5}^{-} 
			+\rho^{2} \omega_{6}^{-}]=0
		\end{aligned}
	\label{rhoquad}
	\end{equation}
	
	Solving Eq. \ref{rhoquad} in Mathematica, retaining the physically acceptable solution (not larger than unity) we have obtained $\rho_{+}$ and $\rho_{-}$ against $\epsilon$.\\
	
	Then meanfield expression for current can be written using Eqs. \ref{apfinal}, \ref{amfinal} and Eqs.\ref{A1} to \ref{A40}, as
	\begin{equation}
		\begin{aligned}
			\text{$J$} = \dfrac{v}{L}\rho_{+}(\rho_{-}-\rho)\{(1-\rho)(1-e^{-(1+\epsilon)/4v}) + \rho(1-e^{-1/4v})\}
		\end{aligned}
	\label{$J$expression3}
	\end{equation}

	In the limit of small and large $\rho$, the solutions for $\rho_{\pm}$ and hence for particle current $J$ take simple forms as we write $a_{\pm}$ in leading order of $\rho$ and $(1-\rho)$ respectively. 
	
	\subsubsection{Small-density approximation} \label{smallrho}
		
	For small $\rho$, we retain leading order terms in $\rho$ in its functions. We obtain,
	\begin{equation}
		a_{+} \approx \rho_{-} [\rho(3\omega_{3}-\omega_{1}-\omega_{4}-2\omega_{5}+\omega_{6})-(\omega_{3}-\omega_{5})] + \rho (\omega_{1}-3\omega_{3}+\omega_{4}) + \omega_{3}
		\label{apsmallrho}
	\end{equation}

	\begin{equation}
		a_{-} \approx \rho_{-} [\rho(3\omega_{3}-\omega_{1}-\omega_{4}-\omega_{5})-\omega_{3}]+\rho(\omega_{1}-3\omega_{3}+\omega_{4}+\omega_{5})+\omega_{3}
		\label{amsmallrho}
	\end{equation}

	Substituting Eqs. \ref{apsmallrho} and \ref{amsmallrho} into Eq. \ref{intermediate} we have,
	\begin{equation}
		\rho_{-} = \dfrac{1-2\omega_{3}}{1-\omega_{3}}\rho
		\label{rhomsmallrho}
	\end{equation}

	\begin{equation}
		\rho_{+} = \dfrac{\rho}{1-\omega_{3}}
		\label{rhopsmallrho}
	\end{equation}
	Scaled current $JL$ takes the form,
	
	\begin{equation}
		JL = \dfrac{v\omega_{5}(1-2\omega_{3})\rho^{2}}{(1-\omega_{3})^2}
		\label{currentsmallrho}
	\end{equation}

	\subsubsection{Large-density approximation} \label{largerho}
	For large $\rho$, we retain terms in leading order of $(1-\rho)$. We obtain,
	\begin{equation}
		a_{+} \approx \omega_{6}(1-\rho) + (\omega_{2}-\omega_{6})(1-\rho)(1-\rho_{-})
		\label{aplargerho}
	\end{equation}
	
	\begin{equation}
		a_{-} \approx \omega_{6}(1-\rho_{-}) + (\omega_{2}+\omega_{5}-2\omega_{6})(1-\rho)(1-\rho_{-})
		\label{amlargerho}
	\end{equation}
	
	Substituting Eqs. \ref{aplargerho} and \ref{amlargerho} into Eq. \ref{intermediate} we have,
	\begin{equation}
		\rho_{-} = 1 - \dfrac{(1-2\omega_{2}-2\omega_{5}+3\omega_{6})}{(1-\omega_{2}-\omega_{5}+\omega_{6})^{2}}(1-\rho)
		\label{rhomlargerho}
	\end{equation}
	
	\begin{equation}
		\rho_{+} = 1-(1-\omega_{6})(1-\rho)
		\label{rhoplargerho}
	\end{equation}
	Scaled current $JL$ takes the form,
	
	\begin{equation}
		JL = \dfrac{v \omega_{6}}{\omega_{6}-1}(1-\rho) 
		\label{currentlargerho}
	\end{equation}.

	\subsection{Simulation results and comparison with mean-field theory} \label{numericsandmean-field}
	
	We present here numerical results obtained from simulation along with analytical calculations from mean-field theory. We have used system size $L=512$ throughout. We have studied variation of density peak $\rho_{+}$, density trough $\rho_{-}$ and particle current $J$ against bulk density $\rho$, interaction strength $\epsilon$ and defect velocity $v$. For $q=0$, it has been observed that magnitude of particle current is maximum around $v=0.16$. Since we are interested in optimum particle transport, $\rho$ and $\epsilon$ dependence of all the quantities have been studied for $v=0.16$. Variation against $\epsilon$ and $v$ have been studied for two different densities, $\rho = 0.29$ and 0.75, well below and above $\rho = 0.5$ while variation against $\rho$ and $v$ are carried out for $\epsilon = 0.6$ and $-0.6$, corresponding to repulsive and attractive interaction respectively.

\begin{figure}[H]
	\centering
	\begin{minipage}{.5\textwidth}
		\centering
		\includegraphics[scale=0.58]{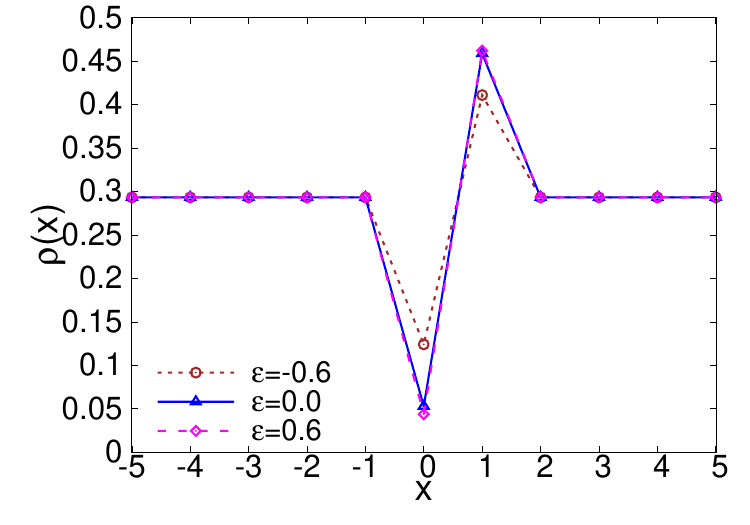}
		\subcaption{$\rho$=0.29}
		\label{dprho0p29}
	\end{minipage}%
	\begin{minipage}{.5\textwidth}	
		\centering
		\includegraphics[scale=0.58]{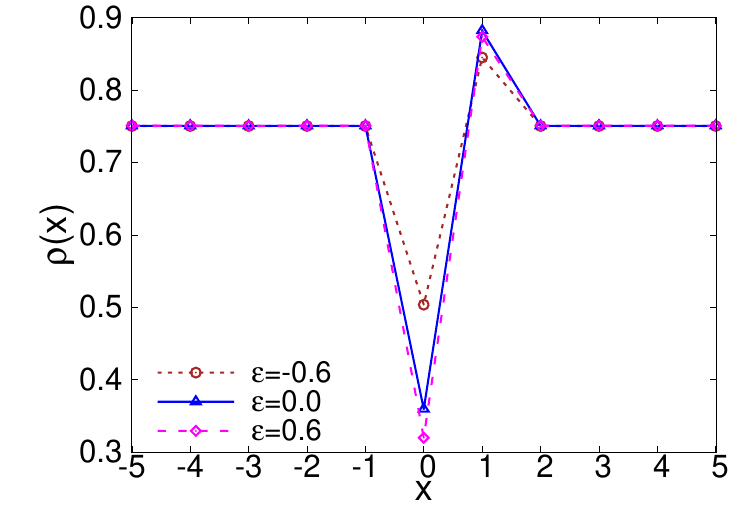}		
		\subcaption{$\rho = 0.75$}
		\label{dprho0p75}
	\end{minipage}%
	\caption{Particle density profile $\rho(x)$ where $x$ denotes the distance from the defect site. For all interactions the defect site has a density trough and its right neighbor has a peak. For  
	attractive interaction the trough and peak are relatively shallower. }
	\label{dpq0}
	\end{figure}

\begin{figure}[H]
		\centering
		\includegraphics[scale=0.63]{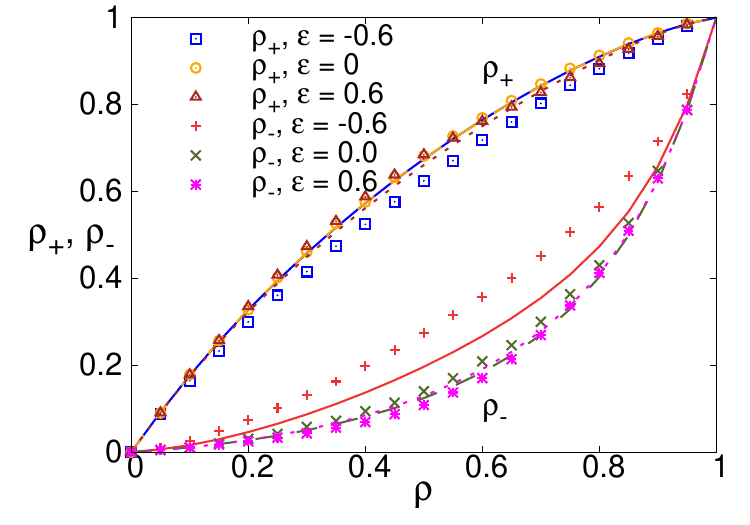}
		\caption{$\rho_{\pm}$ are plotted against bulk density $\rho$. Mean-field results are presented by solid ($\epsilon = -0.6$), dashed ($\epsilon = 0$) and dotted ($\epsilon = 0.6$) lines. $\rho_{+}$ shows a weaker dependence on $\rho$ compared to $\rho_-$ . For attractive interaction $\rho_{+} (\rho_-)$ is noticeably smaller (greater) than that for the hardcore and repulsive interactions. Mean-field results show good agreement for $\epsilon \geq 0$ but  for $\epsilon < 0$ quantitative deviation from numerical data is observed.}
		\label{rhopmvsrhoL}
	\end{figure}

	 From density profiles depicted in Figs. \ref{dprho0p29} and \ref{dprho0p75} and from variation of $\rho_{\pm}$ vs $\rho$ depicted in Fig. \ref{rhopmvsrhoL}, it can be observed that, density peak and trough become more pronounced in case of repulsive interaction. Variation of $\rho_{-}$ with $\rho$ is stronger compared to that of $\rho_{+}$. Such a behavior is also supported by mean-field theory. 

\begin{figure}[H]
		\centering
		\begin{minipage}{.5\textwidth}
			\centering
			\includegraphics[scale=0.58]{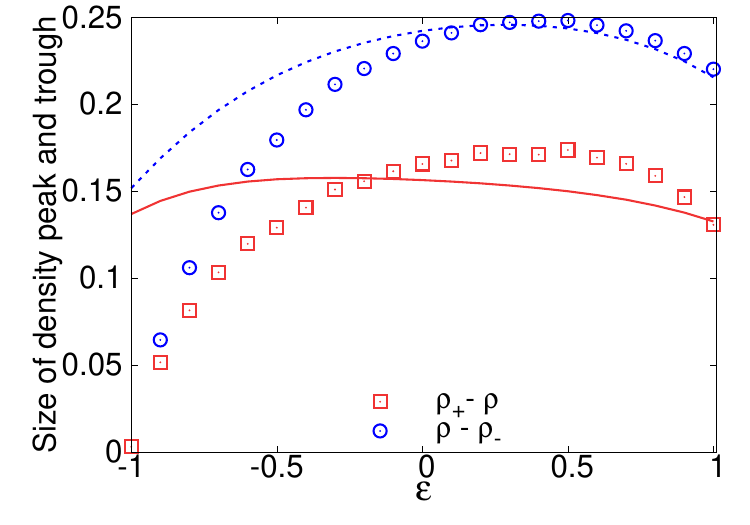}
			\subcaption{$\rho = 0.29$}
			\label{rhopmvsepsrho0p29}
		\end{minipage}%
		\begin{minipage}{.5\textwidth}	
			\centering
			\includegraphics[scale=0.58]{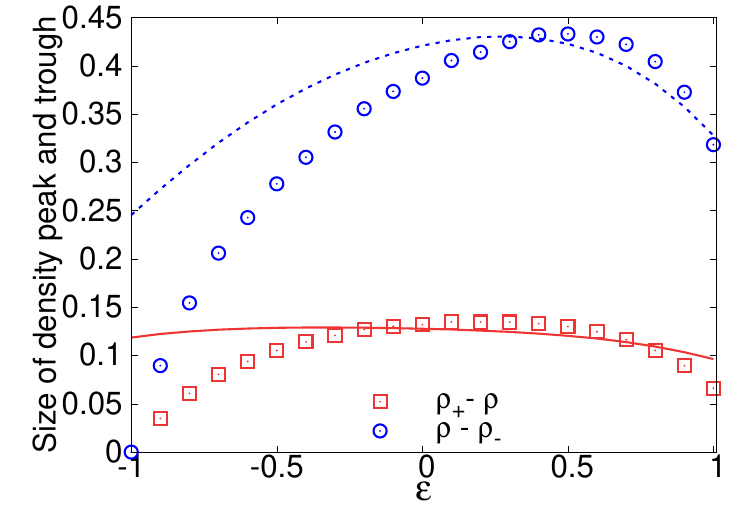}		
			\subcaption{$\rho = 0.75$}
			\label{rhopmvsepsrho0p75}
		\end{minipage}%
		\caption{Size of the density peak and trough, ($\rho_{+}-\rho$) and ($\rho-\rho_{-}$) are plotted against $\epsilon$ in panels (a) and (b). Mean-field results are presented by solid and dotted lines respectively. It is evident from both the panels that depth of the trough is always greater than height of the peak. Non-monotonic variation against $\epsilon$ can be seen in both the quantities which is more pronounced for ($\rho-\rho_{-}$) with a maximum at a large positive $\epsilon$. Mean-field theory can qualitatively capture such behavior in the repulsive region while it fails in the region of attractive interaction.}
		\label{rhopmvseps}
	\end{figure}

	In Fig. \ref{rhopmvseps}, we find that as $\epsilon$ increases from negative to positive values, the differences $(\rho_{+}-\rho)$ and $(\rho-\rho_{-})$ also increase. The variation in both these quantities against $\epsilon$ are non-monotonic with a peak at large positive $\epsilon$ values. Our mean-field calculation captures this non-monotonic behavior but does not provide good quantitative agreement with the numerical data.
	
\begin{figure}[H]
	 	\centering
	 	\begin{minipage}{.5\textwidth}
	 		\centering
	 		\includegraphics[scale=0.58]{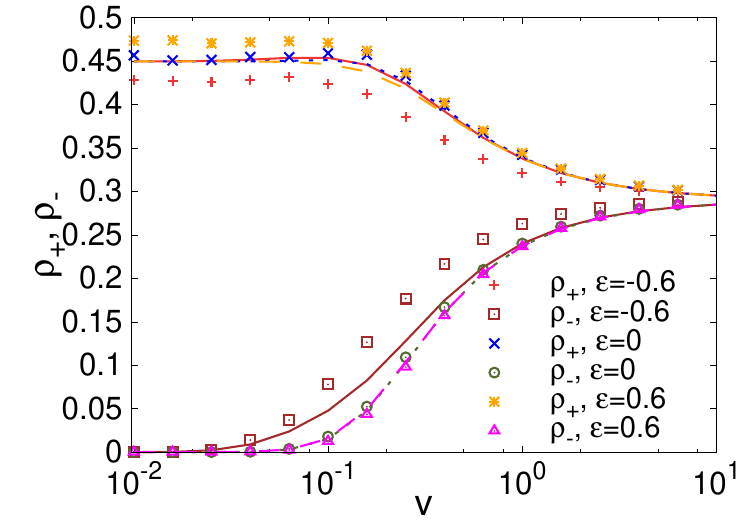}
	 		\subcaption{$\rho = 0.29$}
	 		\label{rhopmvsvrho0p29}
	 	\end{minipage}%
	 	\begin{minipage}{.5\textwidth}	
	 		\centering
	 		\includegraphics[scale=0.58]{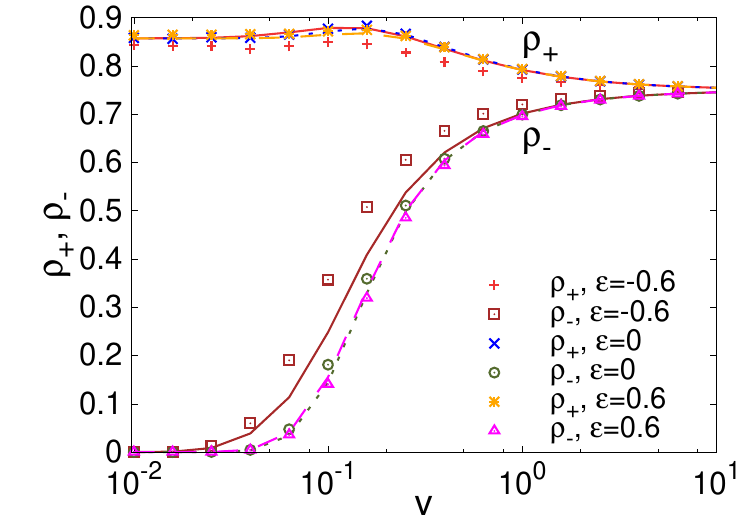}		
	 		\subcaption{$\rho = 0.75$}
	 		\label{rhopmvsvrho0p75}
	 	\end{minipage}%
	 	\caption{$\rho_{\pm}$ are plotted against defect velocity $v$. Mean-field results are presented by solid
	 		($\epsilon = -0.6$), dotted ($\epsilon = 0$) and dashed ($\epsilon = 0.6$) lines which qualitatively capture the variation. For large $v$ both these quantities approach $\rho$ while for small $v$ they show weak variation. Comparing the data for different $\epsilon$ values show that for all $v$ repulsive interaction causes highest (lowest) $\rho_+ (\rho_-)$.}
 			\label{rhopmvsrhoq0}
 	\end{figure}

	 In Figs. \ref{rhopmvsvrho0p29} and \ref{rhopmvsvrho0p75} we show the variation of $\rho_{\pm}$ with $v$ for two different $\rho$ values. For small $v$ a particle can almost always hop out of the defect site but as $v$ increases such a transition may not always be possible because of short residence time of the defect \cite{33}, \cite{34}. Therefore $\rho_{-}$ ($\rho_{+}$) increases (decreases) with $v$, finally saturating to $\rho$ for very large $v$. As $\epsilon$ increases from negative to positive values, for all $v$, $\rho_{-}$ becomes systematically lower and $\rho_{+}$ becomes higher, consistent with what we have shown in Fig. \ref{rhopmvsrhoL}. 
	 
\begin{figure}[H]
		\centering
		\includegraphics[scale=0.58]{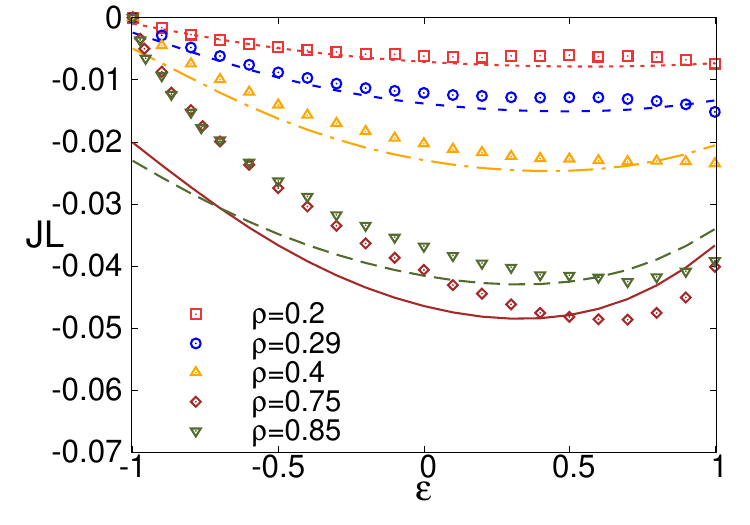}
		\caption{scaled current are plotted against epsilon along with mean-field results (presented by dotted line ($\rho = 0.2$), short-dashed line ($\rho = 0.29$), dot-dashed line ($\rho = 0.4$), solid line ($\rho = 0.75$) and dashed line ($\rho = 0.85$). Current vanishes at $\epsilon=-1$ and remains negative elsewhere. For small and intermediate $\rho$ current is largest for $\epsilon=1$, while for large $\rho$ it shows a peak at a slightly smaller $\epsilon$ value.}
		\label{currentvseps}
	\end{figure}

	In Fig. \ref{currentvseps} the variation of scaled particle current $JL$ with $\epsilon$ has been shown for different $\rho$ values. At $\epsilon = -1$ because of strong attractive interaction among the particles, the system supports one single cluster containing all the particles. Therefore current vanishes in this limit. As $\epsilon$ increases, around the defect site a density profile consisting of peak and trough as shown in Figs. \ref{dprho0p29}, \ref{dprho0p75} is formed and current becomes non-zero. As $\epsilon$ increases further the density peak and trough become more pronounced (as shown in Fig \ref{rhopmvseps}) resulting in larger current magnitude. However, for positive $\epsilon$, current shows qualitatively different variations for small and large $\rho$. For small $\rho$ values, current remains almost constant with $\epsilon$ before showing a mild increase near $\epsilon = 1$. 
     This behavior can be explained from our data in Fig. \ref{rhopmvsepsrho0p29}, where the difference between the two curves (red square and blue diamond), which represents the asymmetry between the sizes of density peak and trough, remains unchanged for a significant range of positive $\epsilon$ and increases when $\epsilon$ is close to $1$. Note that this asymmetry is directly related to the current as shown in Eq. \ref{$J$expression2}. For large $\rho$, on the other hand, current shows a peak at $\epsilon \simeq 0.7$ and decreases beyond that. This behavior is consistent with our data in Fig. \ref{rhopmvsepsrho0p75}, where the two curves are seen farthest apart at that particular $\epsilon$. Note that mean-field theory can qualitatively capture the peak in current for large $\rho$ but for small $\rho$ it is unable to reproduce the upswing shown by our data near $\epsilon = 1$. We find similar disagreement in Fig. \ref{rhopmvsepsrho0p29} as well where mean-field theory hardly captures the variation of trough size. 
     
\begin{figure}[H]
     	\centering
     	\includegraphics[scale=0.58]{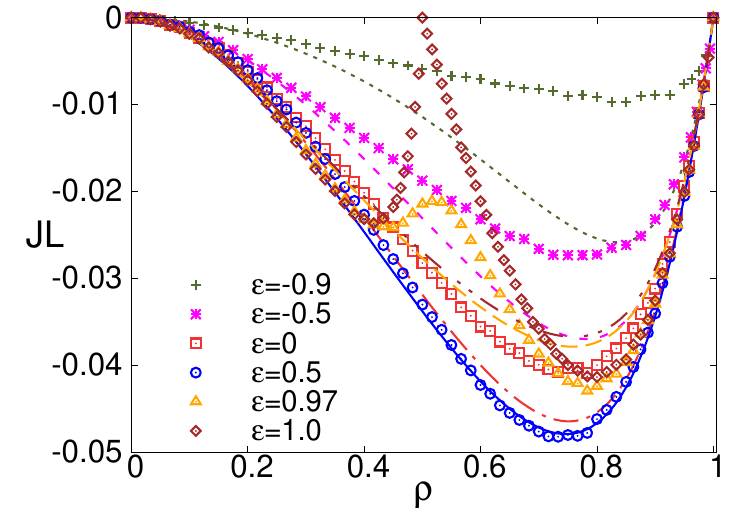}
		\caption{Scaled current vs density for various interaction strengths. Discrete points show simulation data and lines show mean-field calculations. We have used line-styles dotted ($\epsilon=-0.9$), short-dashed ($\epsilon=-0.5$), dashed ($\epsilon=0$), solid ($\epsilon=0.5$), dot-dashed ($\epsilon=0.97$) and a dot-dot-dashed line ($\epsilon=1$). For all epsilon values, current shows non-monotonic variation with density. For attractive interaction and moderate repulsive interaction, current shows a single peak at a density $> 1/2$. However, as repulsion becomes stronger, current shows two peaks, separated by a minimum at $\rho=1/2$. Although mean field theory fails to capture the double peak, we offer an alternative simple explanation in the text.}
			\label{currentvsrho}
	\end{figure}

     Fig. \ref{currentvsrho} shows the plot of scaled current $JL$ vs bulk density $\rho$ for various $\epsilon$ values. In the limit  $\rho \to 0$ and $\rho \to 1$ current vanishes for all $\epsilon$ as expected. We have been able to analytically show (see Eqs. \ref{currentsmallrho}, \ref{currentlargerho}) that in the small density limit current $\sim \rho^2$, while in the large density limit current $\sim (1-\rho)$. This limiting behavior agrees reasonably well with our numerical data. 

    From Fig. \ref{currentvsrho}, we see that, for an intermediate density $\rho^{\ast}$, the current shows a maximum. When $\epsilon$ takes large negative value, the overall magnitude of the current is low because of strong attractive interaction among the particles. As the particle attraction weakens, the current also becomes larger and the peak at $\rho^{\ast}$ gets higher. Our mean-field results successfully capture this trend although $\rho^\ast$ shows dependence on $\epsilon$ unlike a nearly constant $\rho^\ast \simeq 0.75$ obtained from numerics for all $\epsilon$. However, as $\epsilon$ changes sign and becomes positive the repulsive interaction does not favor successive occupied sites. This gives rise to a special point at $\rho=0.5$ and $\epsilon=1$ when the configuration with alternate sites occupied by particles is the only allowed configuration. No transitions are possible from this configuration and hence current vanishes. This is verified from our numerics where current sharply becomes zero at $\rho = 0.5$ for $\epsilon =1$. This generates another peak in current at a lower density $\rho < 0.5$. However, even as $\epsilon$ falls slightly below unity, this effect weakens and the zero of current at half -filled density is replaced by a mild minimum. Unfortunately, our mean-field calculations are unable to capture this effect and predicts a single peak for current for all $\epsilon$.

\begin{figure}[H]
		\centering
		\begin{minipage}{.5\textwidth}
			\centering
			\includegraphics[scale=0.58]{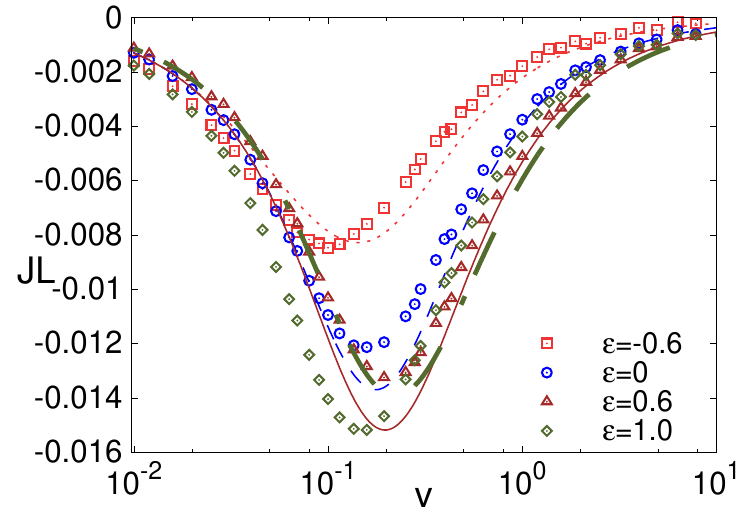}
			\subcaption{$\rho=0.29$}
			\label{Jvsvrho0p29}
		\end{minipage}%
		\begin{minipage}{.5\textwidth}	
			\centering
			\includegraphics[scale=0.58]{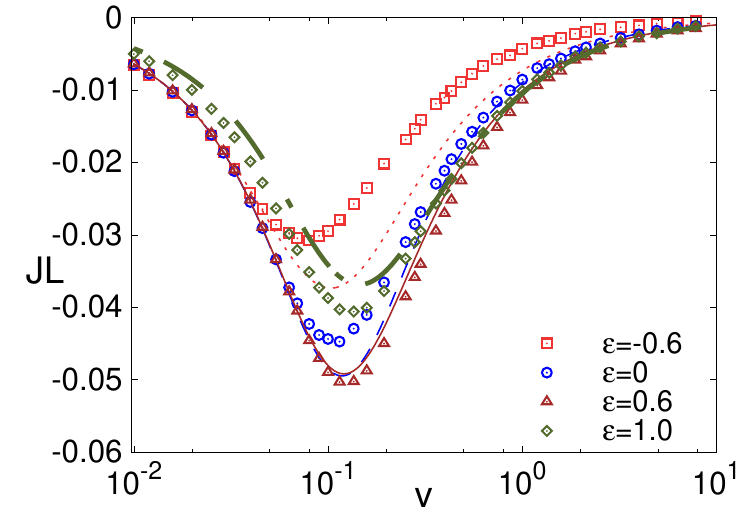}		
			\subcaption{$\rho=0.75$}
			\label{Jvsvrho0p75}
		\end{minipage}%
		\caption{Scaled current $JL$ plotted against defect velocity $v$ along with mean-field results (represented by dotted ($\epsilon = -0.6$), short-dashed ($\epsilon = 0$), solid ($\epsilon = 0.6$) and dot-dashed line ($\epsilon = 1$)) in panels (a)	and (b). Current vanishes in the small $v$ and large $v$ limit and shows a peak in between. The peak height increases as interaction changes from attraction to repulsion. Largest peak is obtained for a large positive value of $\epsilon$. Mean field theory explains the numerical data qualitatively.}
		\label{Jvsv}
	\end{figure}

	Fig. \ref{Jvsv} depicts variation of current with defect velocity $v$ for different interaction strength and two different $\rho$ values. In all cases $v \to 0$ corresponds to the equilibrium limit when current vanishes. For very large $v$ the defect movement becomes too fast for the particles to respond and current vanishes here too. An intermediate $v$ therefore maximizes the current which can be seen both from our numerical data and mean-field calculations. As $\epsilon$ increases from negative to positive values the peak current increases monotonically for $\rho=0.29$ (Fig. \ref{Jvsvrho0p29}), while for $\rho=0.75$ the peak current shows a non-monotonic variation for positive $\epsilon$ (Fig. \ref{Jvsvrho0p75}). This is consistent with the variation observed in Fig. \ref{currentvseps}. 

\begin{figure}[H]
		\begin{minipage}{.5\textwidth}
			\hspace{-3.4 cm}
			\includegraphics[scale=0.23]{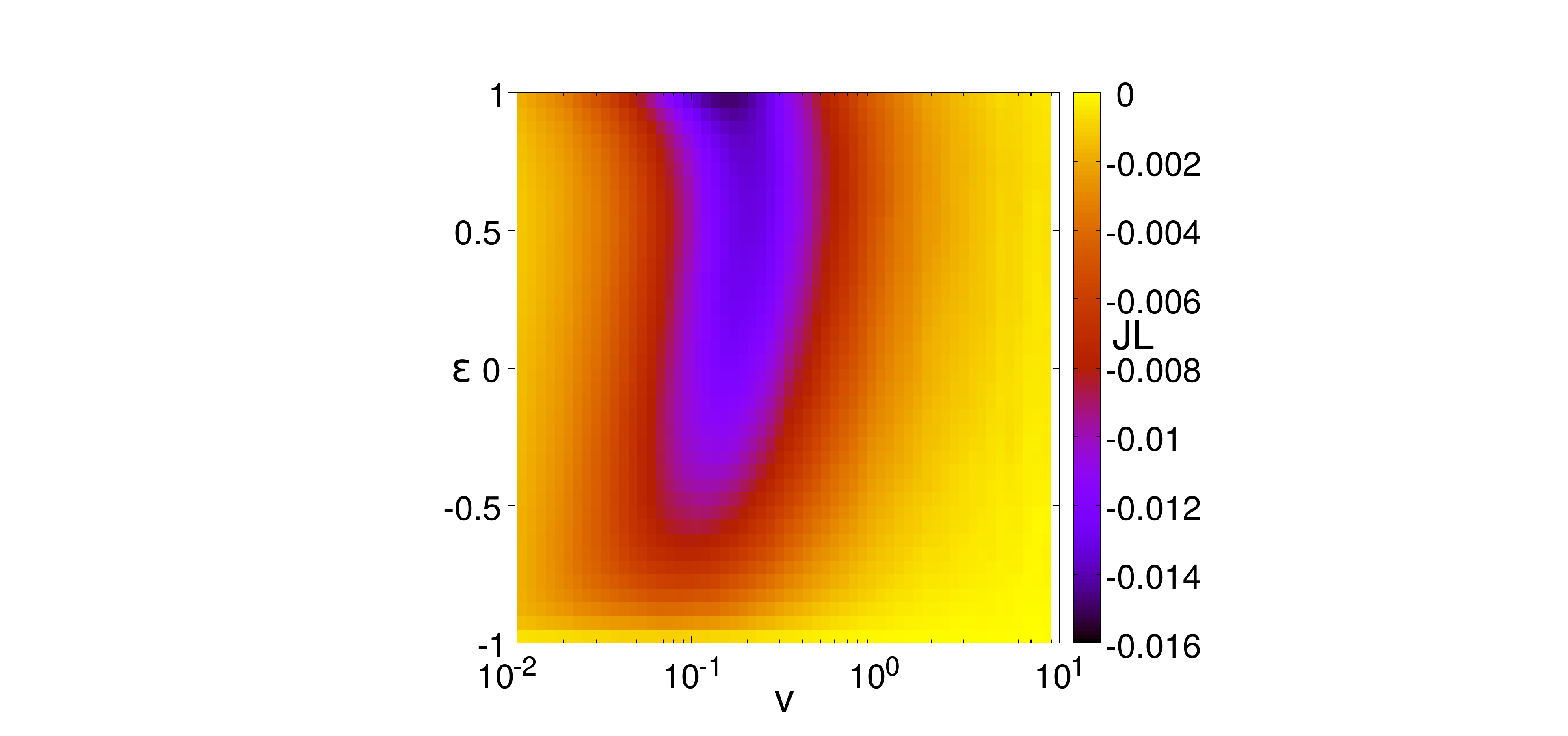}
			\subcaption{$\rho=0.29$}
			\label{heatmap0p29}
		\end{minipage}%
		\begin{minipage}{.5\textwidth}
			\hspace{-3.8 cm}
			\includegraphics[scale=0.23]{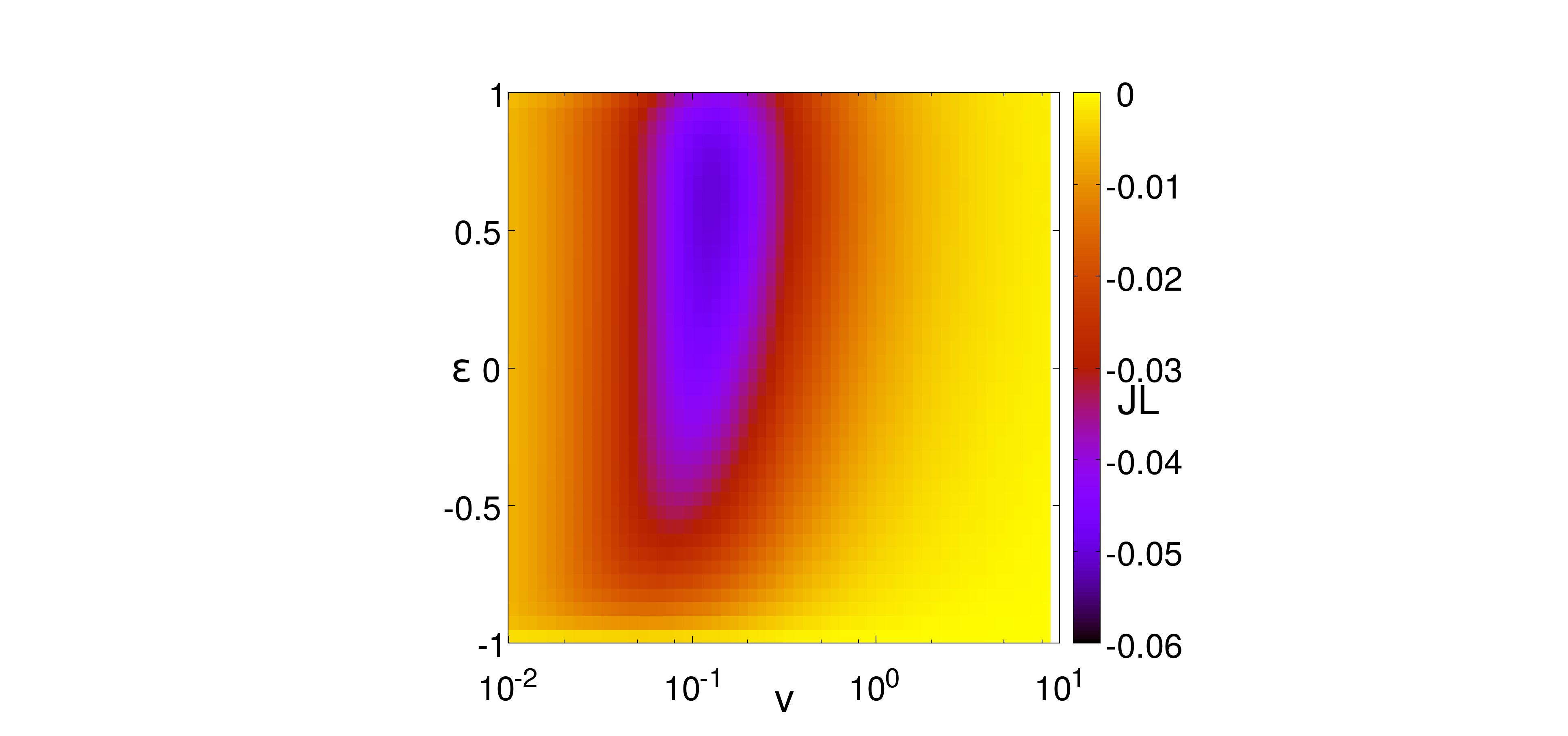}		
			\subcaption{$\rho=0.75$}
			\label{heatmap0p75}
		\end{minipage}
		\begin{minipage}{.5\textwidth}
			\hspace{-3.5 cm}
			\includegraphics[scale=0.23]{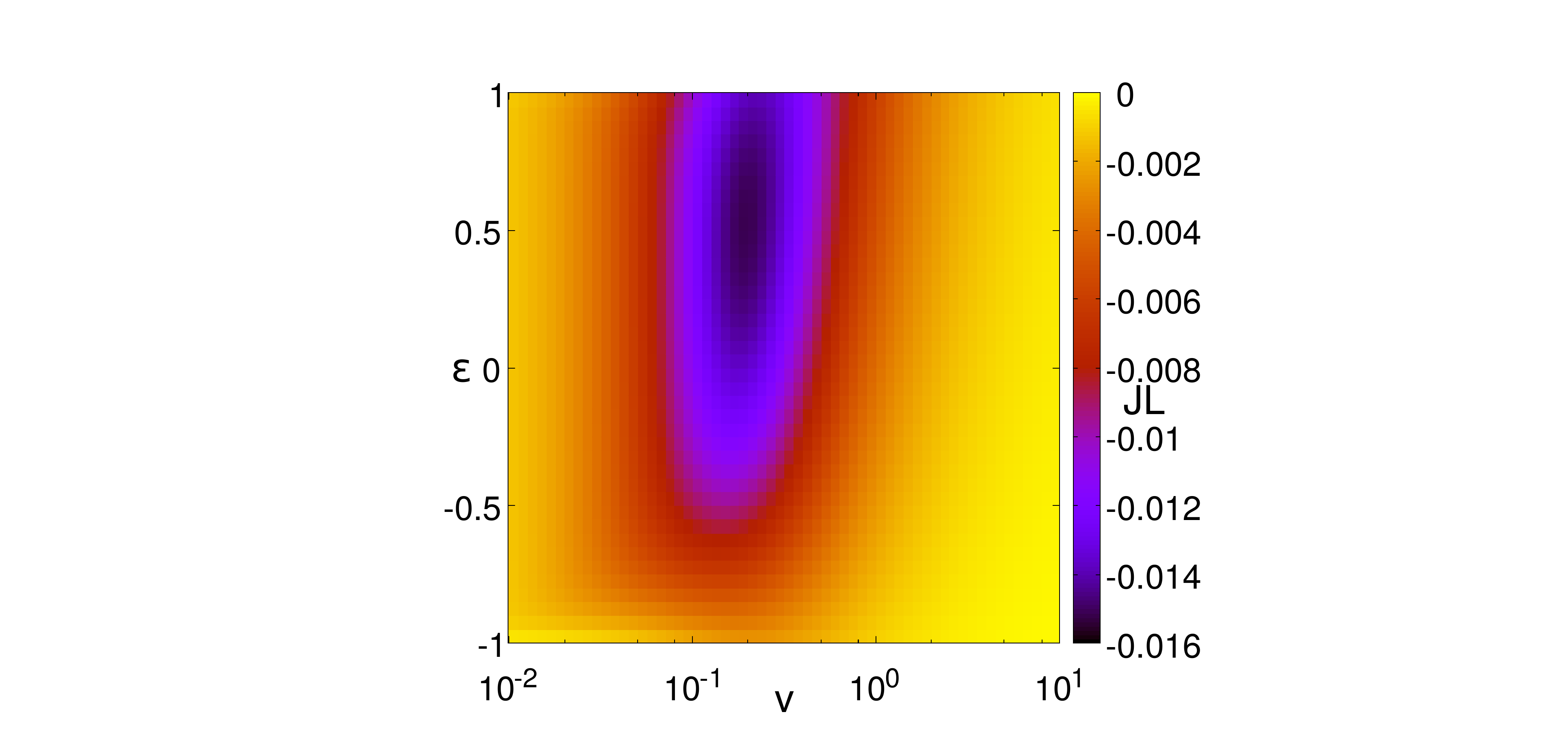}
			\subcaption{$\rho=0.29$}
			\label{heatmapMF0p29}
		\end{minipage}%
		\begin{minipage}{.5\textwidth}
			\hspace{-3.7 cm}
			\includegraphics[scale=0.23]{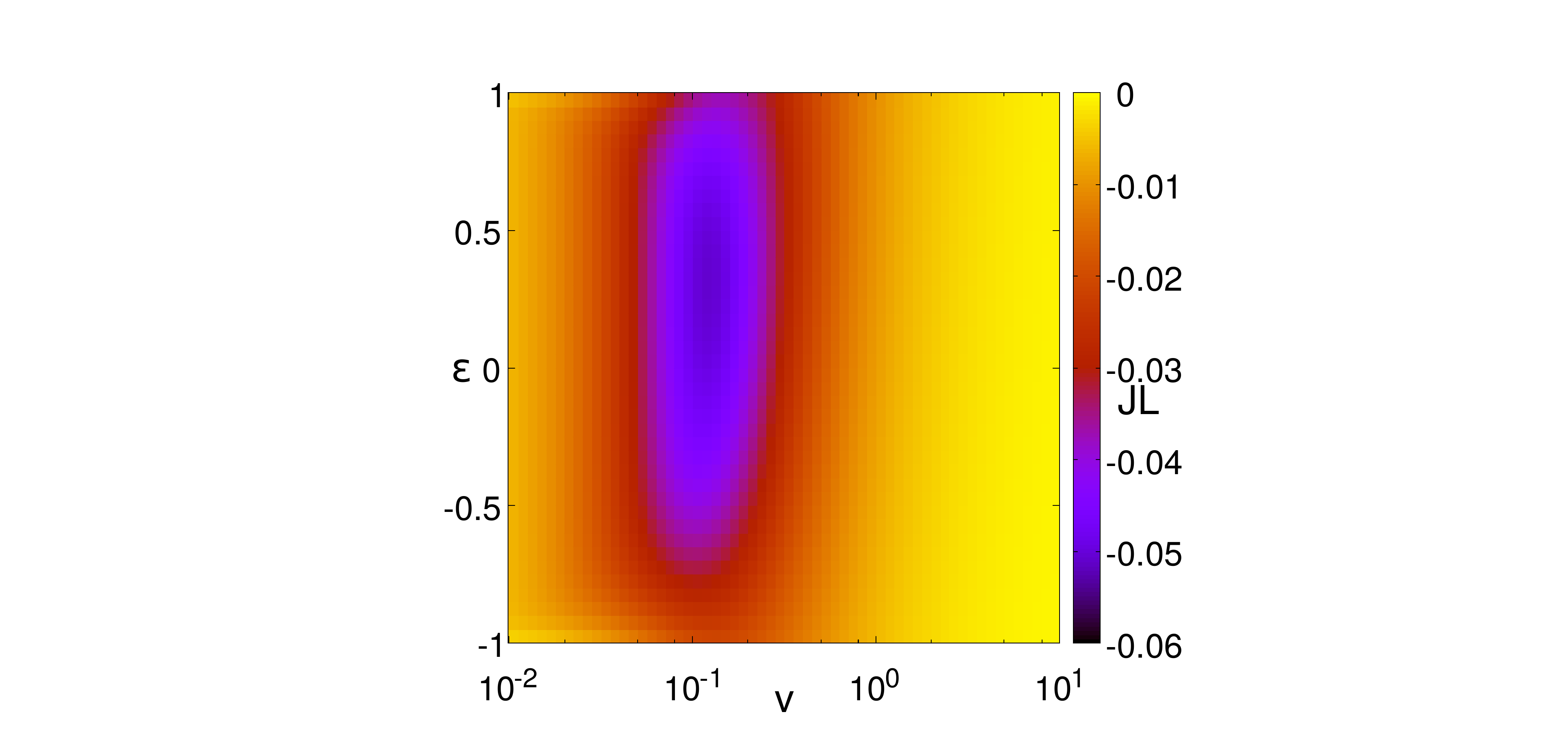}		
			\subcaption{$\rho=0.75$}
			\label{heatmapMF0p75}
		\end{minipage}%
		\caption{Scaled particle current $JL$ is plotted against $\epsilon$ and $v$. Panels (a), (b) represent numerical data while panels (c), (d) show mean-field results. The heat-maps help to trace out the region of $\epsilon$ and $v$ corresponding to the optimum transport in the system. Panel (a) shows that for small density, current is maximum for strongest repulsion $\epsilon=1$, while panel (b) shows that for large $\rho$ a positive $\epsilon < 1$ optimizes the transport. Note however, the scales chosen for left and right panels which show magnitude of the optimum current is much larger when $\rho$ is large. Mean- field results work reasonably well for large $\rho$, but fail to capture the optimum transport regime for small $\rho$.}
		\label{HMq0Jvsveps}
\end{figure}

\begin{figure}[H]
		\begin{minipage}{.5\textwidth}
			\hspace{-3.55 cm}
			\includegraphics[scale=0.23]{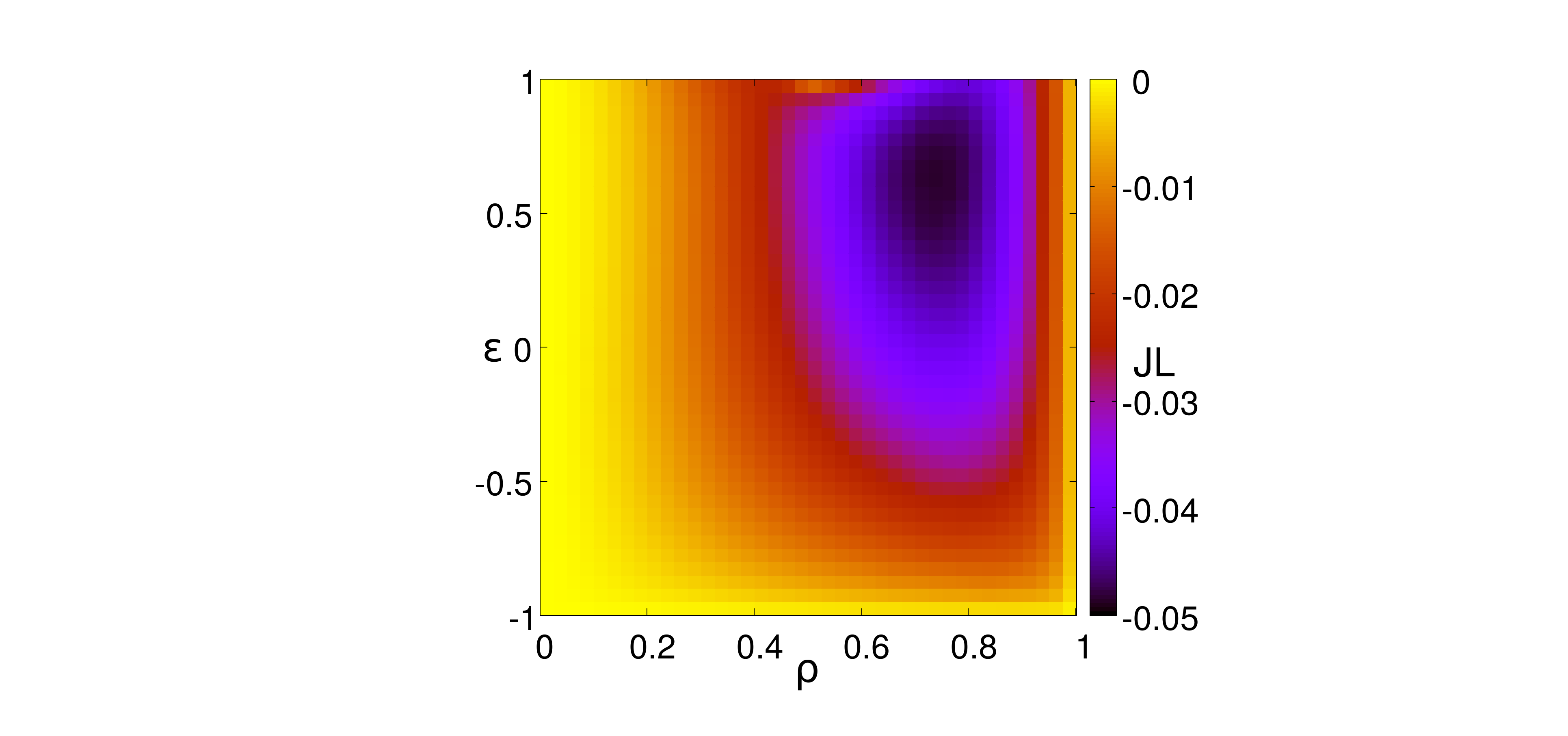}
			\subcaption{$v=0.16$}
			\label{heatmapv0p16}
		\end{minipage}%
		\begin{minipage}{.5\textwidth}
			\hspace{-3.65 cm}
			\includegraphics[scale=0.23]{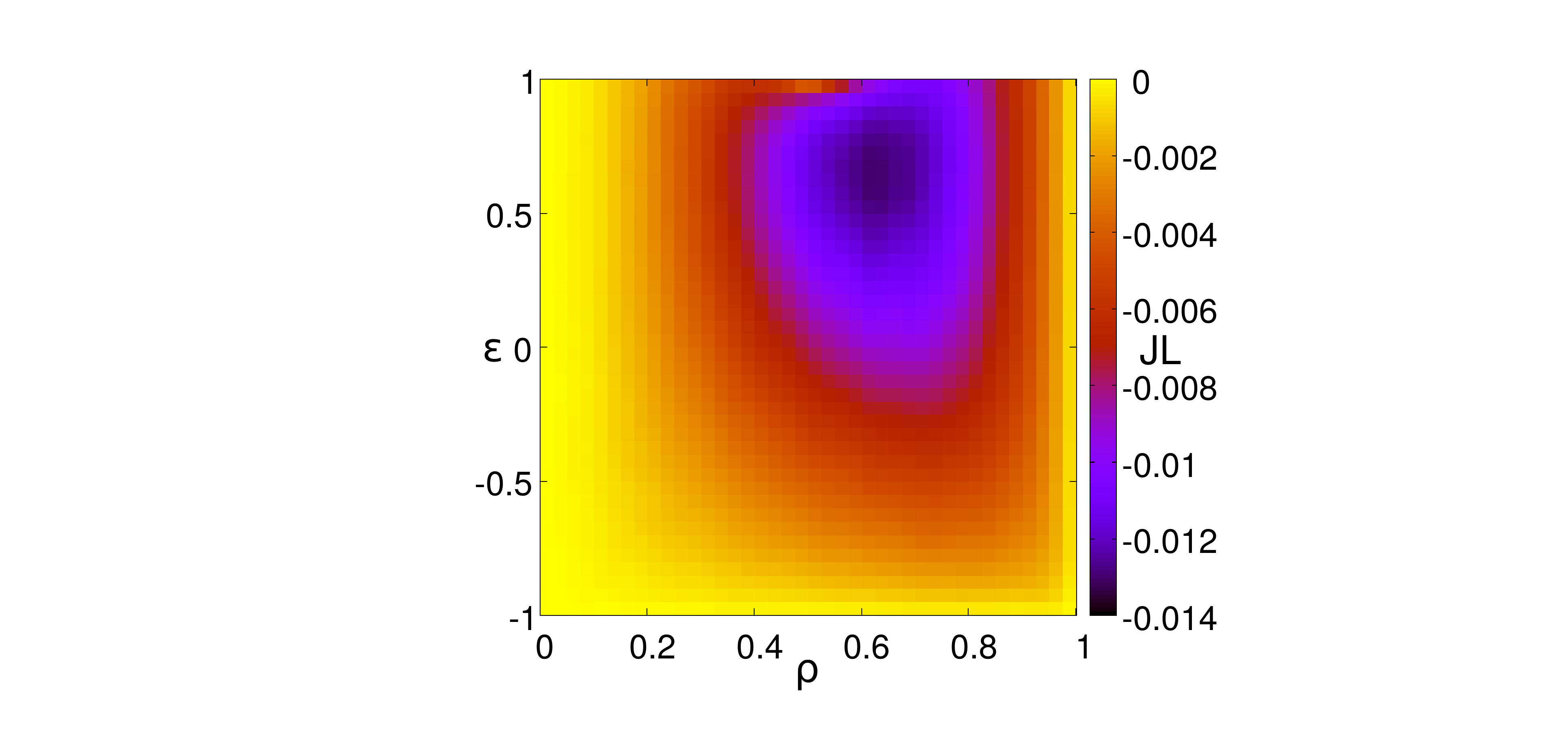}		
			\subcaption{$v=1.0$}
			\label{heatmapv1p0}
		\end{minipage}
		\begin{minipage}{.5\textwidth}
			\hspace{-3.55 cm}
			\includegraphics[scale=0.23]{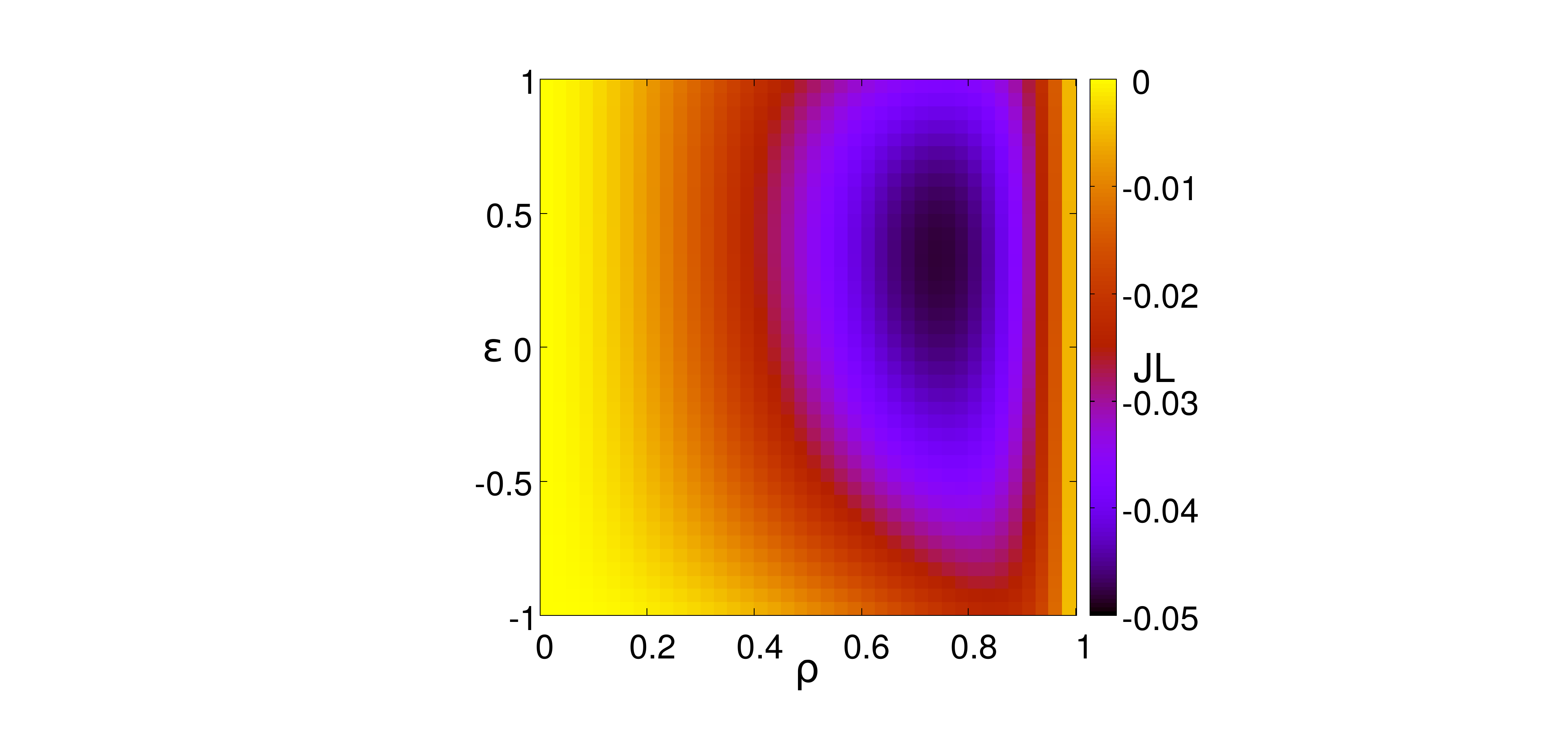}
			\subcaption{$v=0.16$}
			\label{MFheatmapv0p16}
		\end{minipage}%
		\begin{minipage}{.5\textwidth}
			\hspace{-3.65 cm}
			\includegraphics[scale=0.23]{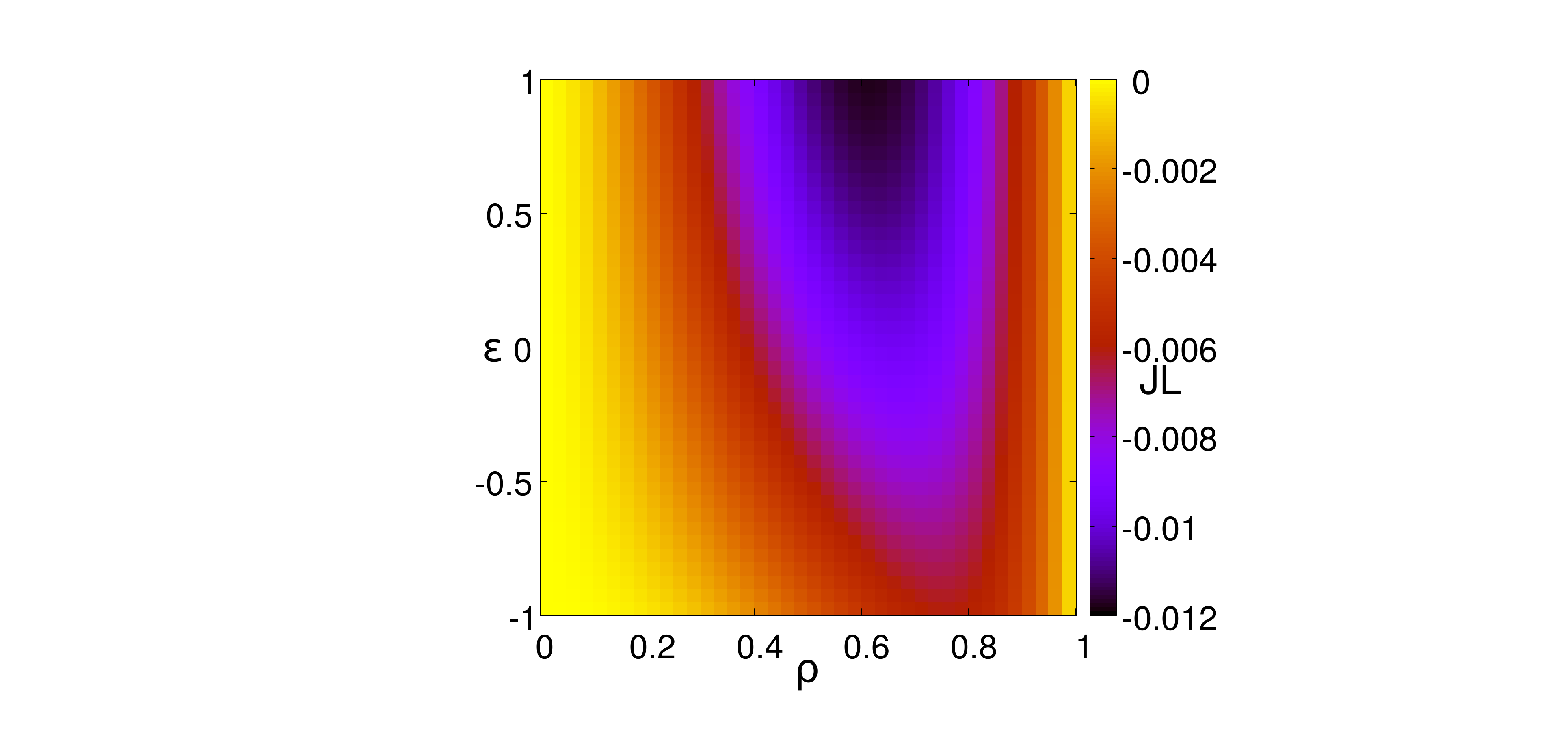}		
			\subcaption{$v=1.0$}
			\label{MFheatmapv1p0}
		\end{minipage}
		\caption{Numerical results for particle current $JL$ is plotted against $\epsilon$ and $\rho$ in panel (a) and (b) while mean-field results are represented in panel (c) and (d). The heat-maps trace out the optimum region for particle current in $\epsilon-\rho$ plane. The region corresponds to large $\rho$ and a high positive $\epsilon$. Magnitude of optimum current is higher in left panels where intermediate $v$ value is used.}
		\label{HMq0Jvsrhoeps}
	\end{figure}

	In Figs. \ref{currentvseps} to \ref{Jvsv}, we have plotted current as a function of one of the three variables $\rho$, $\epsilon$ and $v$, keeping other two constant. To understand the condition of optimum transport, we need to identify how $\rho$, $\epsilon$ and $v$ should be chosen such that the current in the system is maximum. To this end, we present heat-maps in Fig. \ref{HMq0Jvsveps} where we simultaneously vary $\epsilon$ and $v$ for fixed $\rho$. Our numerical data are presented in panels (a), (b) and our mean-field calculations appear in panels (c), (d) in Fig. \ref{HMq0Jvsveps}. These plots clearly show repulsive interaction facilitates particle transport. For smaller density current always increases as $\epsilon$ increases and largest current is obtained at $\epsilon=1$. For larger density on the other hand, very strong repulsion makes certain transitions energetically unfavorable. This hinders particle transport. Therefore in this case optimum transport is obtained at an intermediate $\epsilon$ value. Mean-field calculations manage to reproduce this optimality correctly in Fig. \ref{heatmapMF0p75}, but do not work so well in Fig. \ref{heatmapMF0p29}. Note that the scale used for low density is widely different from that in the high density. This means when the density is low, a condition for optimum transport can be derived but the current is far smaller than optimum regime for high density. This is seen more clearly in Fig. \ref{HMq0Jvsrhoeps}.

 	Fig. \ref{HMq0Jvsrhoeps} depicts the heat-maps for current when $\epsilon$ and $\rho$ are varied keeping $v$ constant. Apart from the usual choice of $v=0.16$, we have also presented data for $v=1$ here. These plots show to obtain the optimum transport, $\rho$ needs to be sufficiently high. Since in the high density regime an intermediate strength of repulsive interaction gives maximum current, the optimum transport happens away from $\epsilon=1$. Note that even in this figure the actual value of maximum current is far higher for $v=0.16$ compared to $v=1$ case.	

	\section{Nonzero bulk-hopping rate: \boldmath$p=1$, $r=0$, $q \neq 0$} \label{non0q}
	
	In the previous section, we had considered the case when the only possible transition in the system is particle hopping out of the defect site. In the present section we consider $q \neq 0$ which allows movement of particles in the bulk of the system. We are interested to find out how this bulk dynamics affects the current. As expected, for very small $q$ our results are similar to what we had presented in the previous section. But as $q$ increases, there is a significant effect on the current. We argue below that the bulk dynamics is expected to make a positive contribution to the current. Note that for $q \neq 0$ the density profile remains homogeneous far from the defect site and therefore the non-vanishing contribution to current comes from the dynamics around the defect site. For small enough $q$ the density profile remains qualitatively similar to our plot in Fig. \ref{dpq0}. A diffusive current will flow between the site with density $\rho_-$ and its left neighbor with density $\rho$. Since $\rho_- < \rho$, this current will be in the positive direction. Therefore, inclusion of bulk dynamics adds a positive component to the system current. We do not have a mean-field theory for this case to support the numerical data. We have considered only small and moderate $q$ in our study.

\begin{figure}[H]
		\begin{minipage}{.5\textwidth}
			\centering
			\includegraphics[scale=0.58]{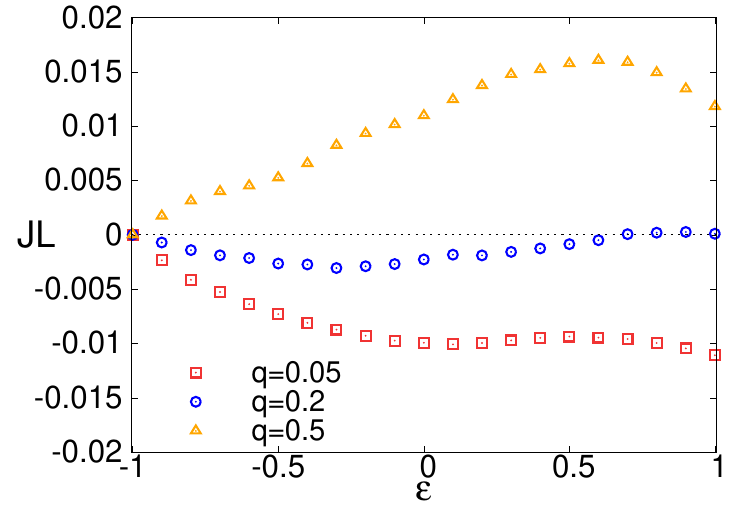}
			\subcaption{$\rho = 0.29$}
			\label{Jvsepsrho0p29qn0}
		\end{minipage}%
		\begin{minipage}{.5\textwidth}	
			\centering
			\includegraphics[scale=0.58]{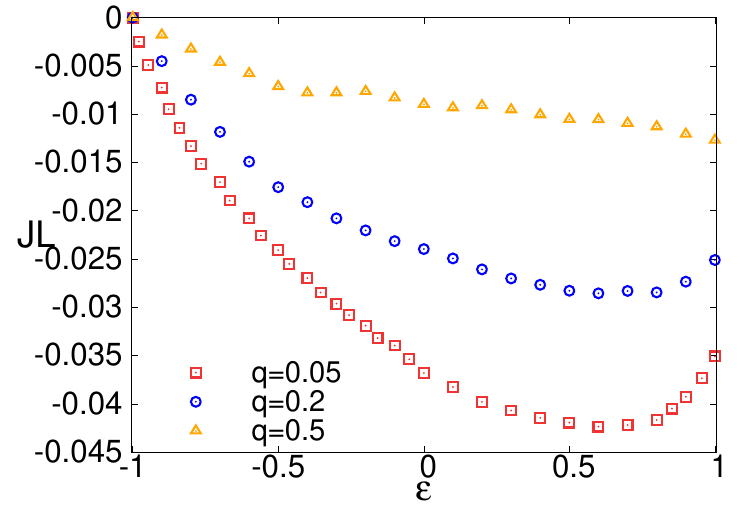}	
			\subcaption{$\rho = 0.75$}
			\label{Jvsepsrho0p75qn0}
		\end{minipage}%
		\caption{Scaled current $JL$ is plotted against epsilon for different $q$ in panel (a) and (b). Current is larger for repulsive interaction compared to attractive ones. At low density and at a moderate $q$, current shows reversal of sign as $\epsilon$ increases while for large density it remains negative for all $q$.}
		\label{Jvsepsnon0q}
	\end{figure}

\begin{figure}[H]
		\centering
		\begin{minipage}{.5\textwidth}
			\centering
			\includegraphics[scale=0.58]{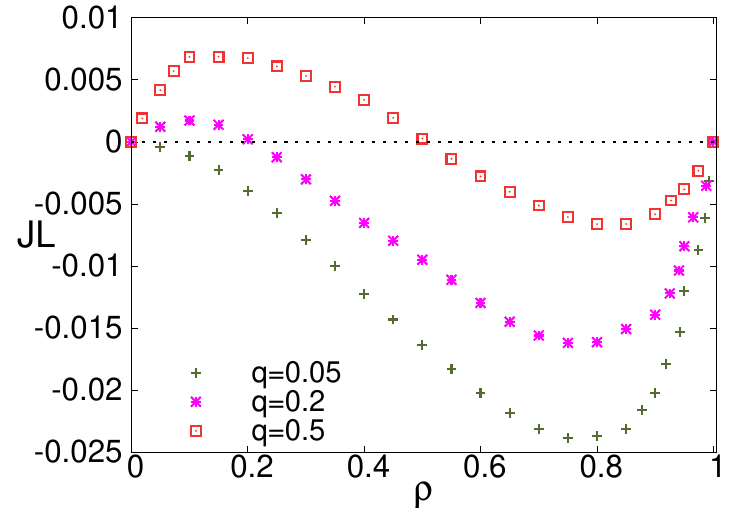}
			\subcaption{$\epsilon=-0.5$}
			\label{Jvsrhoeps-0p5}
		\end{minipage}%
		\begin{minipage}{.5\textwidth}	
			\centering
			\includegraphics[scale=0.58]{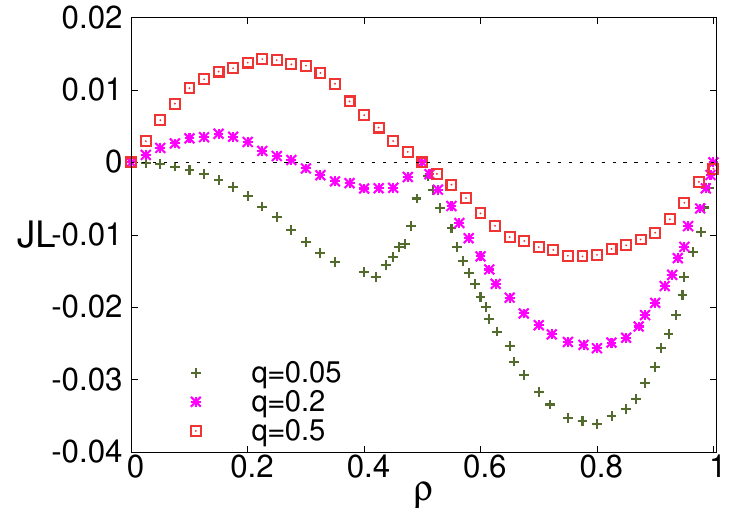}		
			\subcaption{$\epsilon=1.0$}
			\label{Jvsrhoeps1p0}
		\end{minipage}%
		\caption{Variation of scaled current $JL$ with  density $\rho$ for different $q$ values. For small $q$, current shows similar behavior as in $q=0$ case both for attractive and repulsive interaction. For moderate $q$ values current shows a positive peak at small density and a negative peak at large density. For $\epsilon=1$ current crosses zero exactly at $\rho=0.5$.}
		\label{Jvsrhonon0q}
	\end{figure}

\begin{figure}[H]
		\centering
		\begin{minipage}{.5\textwidth}
			\centering
			\hspace{-1 cm}
			\includegraphics[scale=0.58]{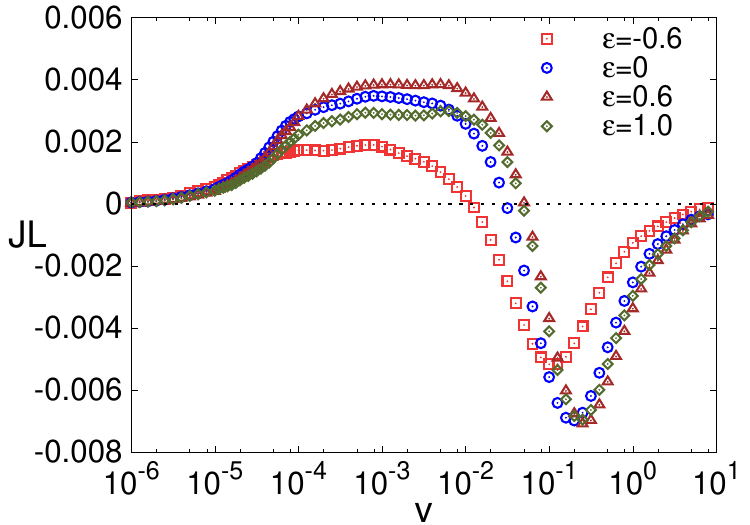}
			\subcaption{$q=0.1$}
			\label{Jvsvrho0p29q0p1}
		\end{minipage}%
		\begin{minipage}{.5\textwidth}	
			\centering
			\hspace{-1 cm}
			\includegraphics[scale=0.58]{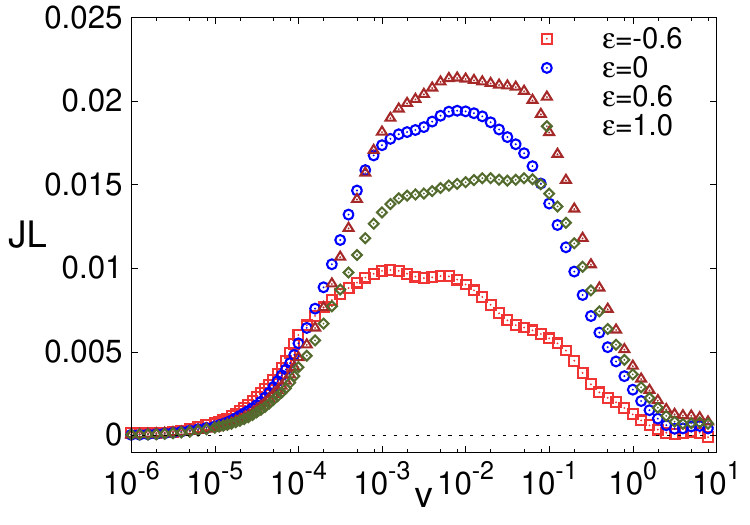}		
			\subcaption{$q=0.5$}
			\label{Jvsvrho0p29q0p5}
		\end{minipage}%
		\caption{Scaled current $JL$ is plotted against defect velocity $v$ for different $\epsilon$. For small $q$ current remains almost flat at small $v$ and reverses its direction at an intermediate $v$, while for moderate $q$ it remains positive throughout for all $\epsilon$.}
		\label{Jvsvnon0q}
	\end{figure}

	In Fig. \ref{Jvsepsnon0q} we plot $JL$ vs $\epsilon$ for fixed $v$ and two different $\rho$ values. For $q=0.05$ the behavior is very similar to the trend observed in Fig. \ref{currentvseps}. As $q$ increases $JL$ becomes more positive as explained above. For small density $JL$ even reverses sign for $q=0.5$. For large density since $JL$ starts from large negative values for $q=0$ it remains negative even when $q=0.5$ although its magnitude decreases because of larger positive contribution coming from bulk dynamics. Similar trends are observed in the variation of $JL$ vs $\rho$ (Fig. \ref{Jvsrhonon0q})or $JL$ vs $v$ (Fig. \ref{Jvsvnon0q}).

\begin{figure}[H]
		\begin{minipage}{.5\textwidth}
			\hspace{-4 cm}
			\includegraphics[scale=0.23]{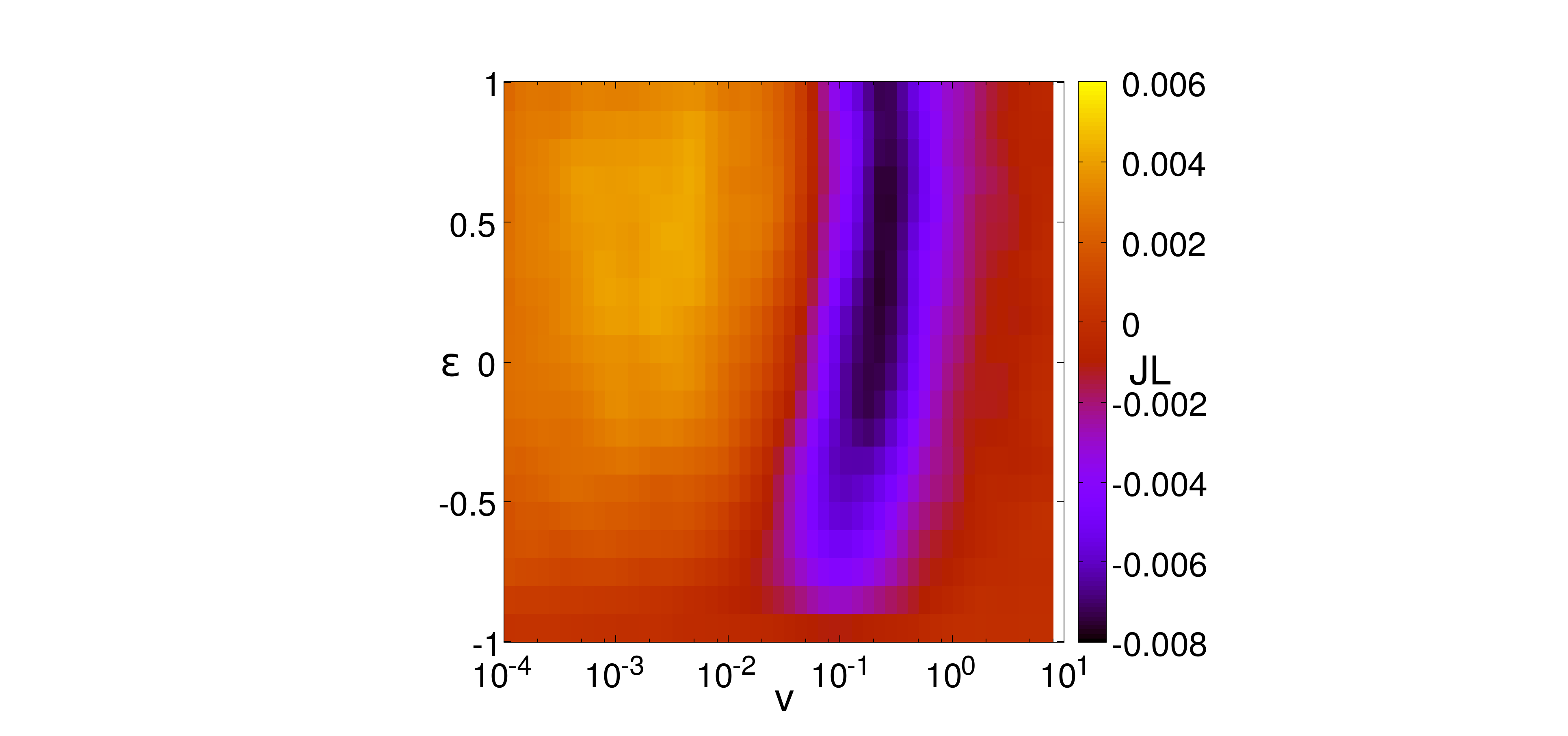}
			\subcaption{$q=0.1$}
			\label{heatmapq0p10p29}
		\end{minipage}%
		\begin{minipage}{.5\textwidth}
			\hspace{-3 cm}
			\includegraphics[scale=0.23]{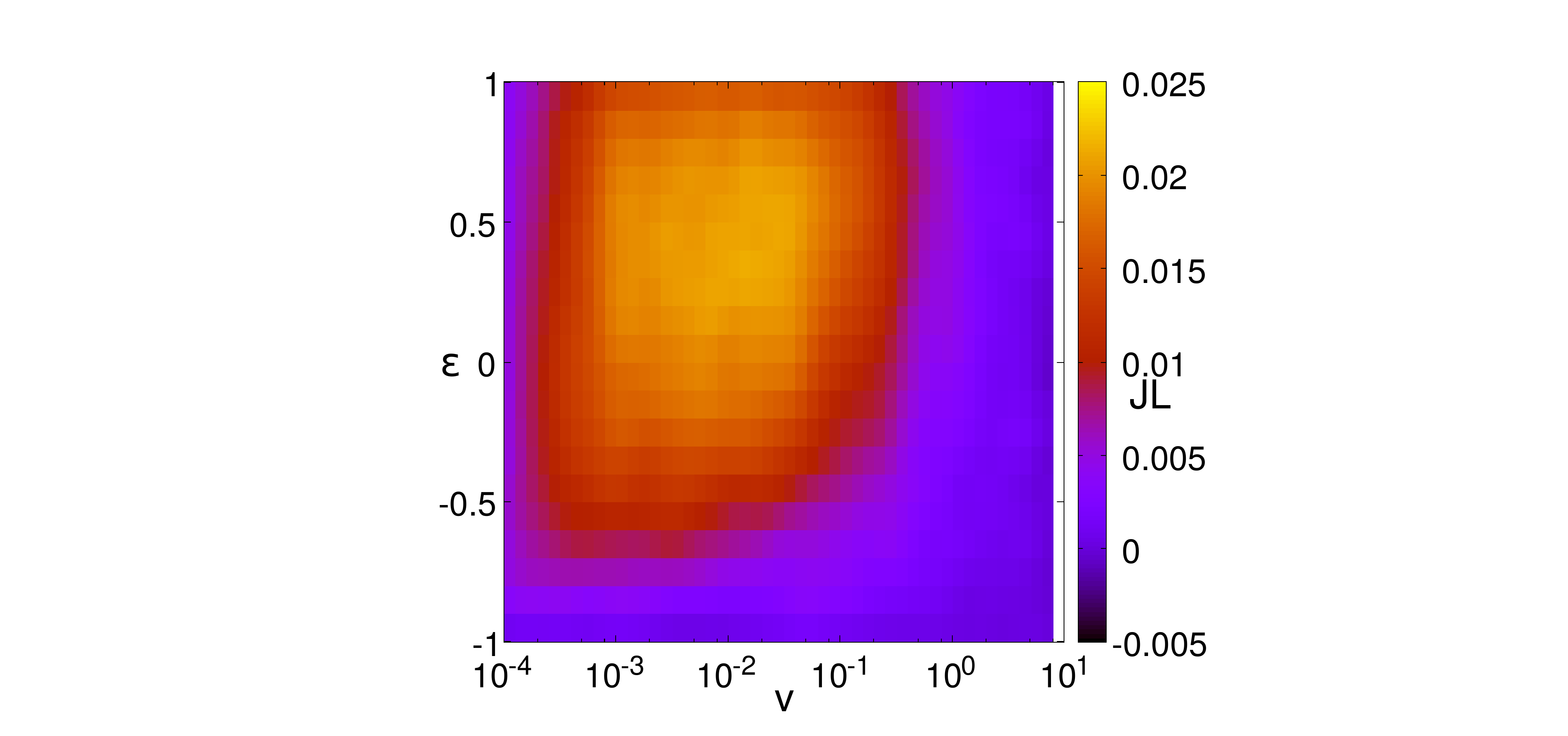}	
			\subcaption{$q=0.5$}
			\label{heatmapq0p50p29}
		\end{minipage}%
		\caption{Numerical results for current $JL$ is plotted against $\epsilon$ and $v$ at $\rho=0.29$. For small $q$, a positive and a negative peak in the variation of current can be observed from panel (a) while panel (b) shows that there exists a single positive peak in its variation. Such peaks occur at large positive $\epsilon$.}
		\label{heatmapJvsvepsn0q}
	\end{figure}

	To identify the parameter regime for optimal transport we show the heatmap in Fig.\ref{heatmapJvsvepsn0q}. We have four relevant parameters here: $\epsilon, v, \rho$ and $q$. For a fixed $\rho$ we show the variation of current in $\epsilon-v$ plane for two different $q$ values. For $q=0.1$ current has both positive and negative peaks, i.e optimum current can flow in the same direction of defect movement or in the opposite direction. This can be clearly seen from Fig. \ref{heatmapq0p10p29}. For $q=0.5$ however, only positive current is possible and optimum transport always happens in the direction of defect movement.

	\section{Summary and concluding remarks}
         \label{summary}
	
	In this paper, we have studied a class of stochastic lattice gases of hardcore particles with nearest-neighbor interaction, where the system is driven by a localized potential barrier (referred to as a ``defect'') moving on a ring. We find that the inter-particle interaction is crucial in controlling particle transport in the system: In the presence of an attractive interaction, the time-averaged dc current decreases, whereas a repulsive interaction increases the current significantly, thus resulting in an interaction-dominated regime of particle transport.
        The moving potential barrier creates a traveling density inhomogeneity, which generates a current in the negative direction, i.e., in the direction opposite to its movement, while the bulk diffusion generates a current in the direction along the barrier movement. As a result, when the bulk hopping (diffusion) rate vanishes, i.e., when $q=0$, the particle current is always negative and shows a negative peak as the barrier movement speed $v$ and bulk density $\rho$ are varied. Quite remarkably, the negative peak in the current is further enhanced when a strong repulsive interaction is present among the particles. On the other hand, for the bulk hopping rate $q \neq 0$, as defect speed $v$, bulk density $\rho$ and interaction strength $\epsilon$ are varied, the particle current shows both positive and negative peaks, which are due to the competition between the positive contribution from bulk diffusion and the negative contribution from the defect movement; however the extent of variation is weaker in this case compared to that for $q=0$. We have been able to identify the precise parameter regime for an optimum transport, which indeed maximizes the magnitude of the current.
        In the case of attractive interaction, a particle prefers to have its nearest neighbor occupied, giving rise to particle clustering. The contribution in the current from the transitions which cause fragmentation of the clusters decreases as the strength of attractive interaction increases, thus resulting in a decreased current. Indeed, unlike repulsive interaction, the current decreases monotonically with the attractive interaction strength, irrespective of defect speed and bulk density.
        To theoretically understand the above results, we perform a modified  mean-field calculation, which - for repulsive interaction, high particle density and negligible bulk diffusion - agrees reasonably well with simulations. Strong attractive interaction causes particle clustering, leading to strong spatial correlations in the system, and the mean-field theory in that case does not work well. Also, for large bulk diffusion, our mean-field theory does not show satisfactory agreement with simulations, again due to the built up of quite strong spatial correlations in the system.

	 The role of inter-particle interactions in controlling particle transport in the presence of a time-periodic drive can be tested in experiments. A periodic potential energy landscape can be created by superimposing external rotating magnetic field on local periodic arrangement of micro-magnets \cite{70}. With the help of this periodic potential, micron-size super paramagnetic beads can be separated from a complex mixture by transporting the beads across a substrate. Then, by tuning the rotational frequency of the external field, the mobility of a specific type of beads can be significantly reduced. the interaction among the paramagnetic colloidal particles can be directly tuned using a modulated ratchet potential \cite{71}. In a system of paramagnetic particles dispersed in water, driven across a striped patterned magnetic garnet film, an external rotating magnetic field induces a periodic potential energy landscape and causes directed motion of the particles. Interestingly, by varying the ellipticity of the rotating magnetic field, the inter-particle interaction can be changed from attractive to repulsive. Our conclusions can be tested in this kind of a setup.

	Throughout this work, we have considered a class of interacting many-particle  models, albeit only on a lattice where particles hop in discrete steps, and it would be quite interesting to investigate particle transport in a continuum. Indeed, in the past, there has already been some progress in this direction where the directed particle transport in continuum was found to be crucially dependent on the precise protocols of the external drive under consideration. For example, a sinusoidally varying traveling wave potential is known to generate a current always in the direction of the traveling wave for a system of particles diffusing on a one-dimensional ring \cite{200}.
        However, in a previous work, from our group it was demonstrated, using numerical simulations, that a moving potential barrier can in fact generate current in either direction, depending on whether the potential moves uniformly or in discrete jumps \cite{201}.
Interestingly, some recent studies have reported multiple current reversal for Brownian particles in the presence of a traveling wave potential \cite{202, 203, 204}.
In a slightly different context, Ref. \cite{72} numerically investigated the effect of interaction on particle transport in asymmetric channels and observed that, depending on the frequency of the external periodic drive, it is possible to enhance transport by tuning the interaction potential.
For single-file diffusion of colloidal particles in an external time-varying force field, various types of interactions such as Weeks-Chandler-Andersen, Yukawa, and super paramagnetic potentials were considered \cite{73}, and anomalous transport was observed.
Indeed, theoretical understanding of transport in continuum, with such realistic potentials and in the presence of a time-periodic drive, will be of significant interest in the context of obtaining the most efficient directed flow.
       However, it is worth mentioning here that analytical calculations in such a many-particle continuum models is quite challenging. In this scenario, theoretical studies of lattice models such as those presented here are quite relevant and useful, particularly in terms of analytically calculating the transport properties of these systems, and could initiate further research in this direction.

	\appendix
	
	\renewcommand{\theequation}{A-\arabic{equation}}
	\setcounter{equation}{0}  
	\section*{Appendix : Calculation of \boldmath$a_{\pm}$ for \boldmath$q = r = 0$}
	
	In the main text, $a_{+}(a_{-})$ is defined as the conditional probability that given the defect site is occupied, a particle hops from the defect site to its empty right (left) neighbor site during the residence time $\tau$ of the defect at a single site. Eq. \eqref{apmexact}  provides a formal mathematical definition for $a_\pm$. In this appendix we outline the calculation for $\omega^{\pm}_{i}$ with $i=1,2...,6$ as explicit functions of $\epsilon$ and $v$.

	Let $\hat{1}$ ($\hat{0}$) denote an occupied (empty) defect site. When a particle hops rightward from the defect site, there are six possible local configurations, which are : $00\hat{1}01$, $10\hat{1}01$, $00\hat{1}00$, $10\hat{1}00$, $1\hat{1}00$, $1\hat{1}01$. We number them as $i=1,2,...,6$. Similarly, for leftward hopping possibilities are: $10\hat{1}00$, $10\hat{1}01$, $00\hat{1}00$, $00\hat{1}01$, $00\hat{1}1$ and $10\hat{1}1$. For a system of size $L$ we divide one Monte Carlo step in $L$ time-intervals of length $dt=1/L$, where $L \gg 1$. For a specific local configuration $i$,  $\omega^{+}_{i} (\omega^{-}_{i})$ is defined as the probability that a particle hops from the defect site to its right (left) neighboring site during $\tau$. The configurations are numbered in such a way, that $\omega^+_i = \omega^-_i = \omega_i$. Below we discuss only the rightward hopping events, which can be easily generalized for leftward hopping as well. 
	
	\subsection*{Calculation for $\omega_{1}=\omega(00\hat{1}01)$}

	For the local configuration $00\hat{1}01$ the probability that the particle hopping event takes place during the first infinitesimal time step $dt$ is given by $(1-\epsilon)pdt/4$ (see Fig. \ref{TheModelFig}). Probability that no hopping takes place in this interval is 
	\begin{equation} 
		\biggl(1 - \dfrac{pdt(1-\epsilon)}{4} - \dfrac{pdt}{4}\biggr) \label{A1} \end{equation}
	which includes the possibilities both leftward and rightward hopping attempts was unsuccessful. The probability that the hopping event takes place after time $2dt$ is therefore,
	\begin{equation}
		\biggl(1 - \dfrac{pdt(1-\epsilon)}{4} - \dfrac{pdt}{4}\biggr)\biggl (\dfrac{(1-\epsilon)pdt}{4}\biggr ).
		\label{A2}
	\end{equation}
	Similarly the probability that it takes place at time $3dt$ is 
	\begin{equation}
		\biggl(1 - \dfrac{pdt(1-\epsilon)}{4} - \dfrac{pdt}{4}\biggr)^{2}\biggl (\dfrac{(1-\epsilon)pdt}{4}\biggr )
		\label{A3}
	\end{equation}
	and so on. So the probability $\omega_{1}$ that the hopping happens in any of the $\tau/dt$ time steps is
	\begin{equation*}
		\dfrac{(1-\epsilon)pdt}{4} \biggl[1 + \biggl(1 - \dfrac{(2-\epsilon)pdt}{4}\biggr ) + \biggl (1 - \dfrac{(2-\epsilon)pdt}{4}\biggr )^2 + ..... + \biggl (1 - \dfrac{(2-\epsilon)pdt}{4}\biggr )^{(\tau/dt)-1} \biggr ) \biggr]
	\end{equation*}
	\begin{align}
		= \dfrac{(1-\epsilon)pdt}{4}     \biggl(\dfrac{1-(1-(2-\epsilon)pdt/4)^{\tau/dt}}{1-(1-(2-\epsilon)pdt/4)}\biggr) = \dfrac{1-\epsilon}{2-\epsilon}\biggl(1-e^{-(2-\epsilon)/4v}\biggr)
		\label{A4}
	\end{align}
	where we have used $\tau = 1/v$ and $dt \to 0$.

	\subsection*{Results for remaining $\omega$}
	
	Following similar steps as outlined above, expressions for all other $\omega$ can be derived. We directly present the final results here
	
	\bea
	\nonumber
	\omega_{2}=\omega(10\hat{1}01) &=& \frac{1}{2}\biggl(1-e^{-p(1-\epsilon)/2v}\biggr) \\
	\nonumber
	\omega_{3}=\omega(00\hat{1}00) &=& \dfrac{1}{2}\biggl(1-e^{-p/2v}\biggr) \\
	\omega_{4}=\omega(10\hat{1}00) &=& \dfrac{1}{2-\epsilon}\biggl(1-e^{-p(2-\epsilon)/4v}\biggr) \\
	\nonumber
	\omega_{5}=\omega(1\hat{1}00) &=&  \biggl(1-e^{-p(1+\epsilon)/4v}\biggr) \\
	\nonumber
	\omega_{6}=\omega(1\hat{1}01) &=& \biggl(1-e^{-p/4v}\biggr) 
	\eea

	\subsection*{Expressions for ${\mathcal C}_i^\pm$ }
	
	We provide the formal definitions for  ${\mathcal C}_i^\pm$ below. These denote the conditional probability of a specific local configuration, given that the defect site is occupied.

	
	\begin{equation}
		{\mathcal C}_{1}^{+} = \text{Prob}.(00\hat{1}01|\hat{1}) = \dfrac{\bigg\langle(1-\eta^{(\alpha)}_{\alpha - 1})(1-\eta^{(\alpha)}_{\alpha})\eta^{(\alpha)}_{\alpha + 1}(1-\eta^{(\alpha)}_{\alpha + 2})\eta^{(\alpha)}_{\alpha + 3}\bigg\rangle}{\langle\eta^{(\alpha)}_{\alpha + 1}\rangle} 
		\label{A29}
	\end{equation}
	
	\begin{equation}
		{\mathcal C}_{2}^{+} = \text{Prob}.(10\hat{1}01|\hat{1}) = \dfrac{\bigg\langle\eta^{(\alpha)}_{\alpha - 1}(1-\eta^{(\alpha)}_{\alpha})\eta^{(\alpha)}_{\alpha + 1}(1-\eta^{(\alpha)}_{\alpha + 2})\eta^{(\alpha)}_{\alpha + 3}\bigg\rangle}{\langle\eta^{(\alpha)}_{\alpha + 1}\rangle} 
		\label{A30}
	\end{equation}
	
	\begin{equation}
		{\mathcal C}_{3}^{+} = \text{Prob}.(00\hat{1}00|\hat{1}) = \dfrac{\bigg\langle(1-\eta^{(\alpha)}_{\alpha - 1})(1-\eta^{(\alpha)}_{\alpha})\eta^{(\alpha)}_{\alpha + 1}(1-\eta^{(\alpha)}_{\alpha + 2})(1-\eta^{(\alpha)}_{\alpha + 3})\bigg\rangle}{\langle\eta^{(\alpha)}_{\alpha + 1}\rangle} 
		\label{A31}
	\end{equation}
	
	\begin{equation}
		{\mathcal C}_{4}^{+} = \text{Prob}.(10\hat{1}00|\hat{1}) = \dfrac{\bigg\langle\eta^{(\alpha)}_{\alpha - 1}(1-\eta^{(\alpha)}_{\alpha})\eta^{(\alpha)}_{\alpha + 1}(1-\eta^{(\alpha)}_{\alpha + 2})(1-\eta^{(\alpha)}_{\alpha + 3})\bigg\rangle}{\langle\eta^{(\alpha)}_{\alpha + 1}\rangle} 
		\label{A32}
	\end{equation}
	
	\begin{equation}
		{\mathcal C}_{5}^{+} = \text{Prob}.(1\hat{1}00|\hat{1}) = \dfrac{\bigg\langle\eta^{(\alpha)}_{\alpha}\eta^{(\alpha)}_{\alpha + 1}(1-\eta^{(\alpha)}_{\alpha + 2})(1-\eta^{(\alpha)}_{\alpha + 3})\bigg\rangle}{\langle\eta^{(\alpha)}_{\alpha + 1}\rangle} 
		\label{A33}
	\end{equation}
	
	\begin{equation}
		{\mathcal C}_{6}^{+} = \text{Prob}.(1\hat{1}01|\hat{1}) = \dfrac{\bigg\langle\eta^{(\alpha)}_{\alpha}\eta^{(\alpha)}_{\alpha + 1}(1-\eta^{(\alpha)}_{\alpha + 2})\eta^{(\alpha)}_{\alpha + 3}\bigg\rangle}{\langle\eta^{(\alpha)}_{\alpha + 1}\rangle} 
		\label{A34}
	\end{equation}
	
	\begin{equation}
		{\mathcal C}_{1}^{-} = \text{Prob}.(10\hat{1}00|\hat{1}) = \dfrac{\bigg\langle\eta^{(\alpha)}_{\alpha - 1}(1-\eta^{(\alpha)}_{\alpha})\eta^{(\alpha)}_{\alpha + 1}(1-\eta^{(\alpha)}_{\alpha + 2})(1-\eta^{(\alpha)}_{\alpha + 3})\bigg\rangle}{\langle\eta^{(\alpha)}_{\alpha + 1}\rangle}
		\label{A35}
	\end{equation}
	
	\begin{equation}
		{\mathcal C}_{2}^{-} = \text{Prob}.(10\hat{1}01|\hat{1}) = \dfrac{\bigg\langle\eta^{(\alpha)}_{\alpha - 1}(1-\eta^{(\alpha)}_{\alpha})\eta^{(\alpha)}_{\alpha + 1}(1-\eta^{(\alpha)}_{\alpha + 2})\eta^{(\alpha)}_{\alpha + 3}\bigg\rangle}{\langle\eta^{(\alpha)}_{\alpha + 1}\rangle}
		\label{A36}
	\end{equation}
	
	\begin{equation}
		{\mathcal C}_{3}^{-} = \text{Prob}.(00\hat{1}00|\hat{1}) = \dfrac{\bigg\langle(1-\eta^{(\alpha)}_{\alpha - 1})(1-\eta^{(\alpha)}_{\alpha})\eta^{(\alpha)}_{\alpha + 1}(1-\eta^{(\alpha)}_{\alpha + 2})(1-\eta^{(\alpha)}_{\alpha + 3})\bigg\rangle}{\langle\eta^{(\alpha)}_{\alpha + 1}\rangle}
		\label{A37}
	\end{equation}
	
	\begin{equation}
		{\mathcal C}_{4}^{-} = \text{Prob}.(00\hat{1}01|\hat{1}) = \dfrac{\bigg\langle(1-\eta^{(\alpha)}_{\alpha - 1})(1-\eta^{(\alpha)}_{\alpha})\eta^{(\alpha)}_{\alpha + 1}(1-\eta^{(\alpha)}_{\alpha + 2})\eta^{(\alpha)}_{\alpha + 3}\bigg\rangle}{\langle\eta^{(\alpha)}_{\alpha + 1}\rangle}
		\label{A38}
	\end{equation}
	
	\begin{equation}
		{\mathcal C}_{5}^{-} = \text{Prob}.(00\hat{1}1|\hat{1}) = \dfrac{\bigg\langle(1-\eta^{(\alpha)}_{\alpha - 1})(1-\eta^{(\alpha)}_{\alpha})\eta^{(\alpha)}_{\alpha + 1}\eta^{(\alpha)}_{\alpha + 2}\bigg\rangle}{\langle\eta^{(\alpha)}_{\alpha + 1}\rangle}
		\label{A39}
	\end{equation}
	
	\begin{equation}
		{\mathcal C}_{6}^{-} = \text{Prob}.(10\hat{1}1|\hat{1}) = \dfrac{\bigg\langle\eta^{(\alpha)}_{\alpha - 1}(1-\eta^{(\alpha)}_{\alpha})\eta^{(\alpha)}_{\alpha + 1}\eta^{(\alpha)}_{\alpha + 2}\bigg\rangle}{\langle\eta^{(\alpha)}_{\alpha + 1}\rangle}
		\label{A40}
	\end{equation}

\end{document}